\documentclass{bcp}

\usepackage{tocbibind}

\def\LaTeX{L\kern -.36em\raise .3ex\hbox{\sc a}\kern -.15em T\kern -.1667em%
\lower .7ex\hbox{E}\kern -.125em X}

\def\refname{References}
\def\thebibliography#1{\vskip24pt plus4pt minus4pt
\small
\centerline{\bf \refname}
\vglue11pt plus2pt minus2pt
\nobreak\list
 {[\arabic{enumi}]}{\settowidth\labelwidth{[#1]}\leftmargin\labelwidth
 \parsep-3pt plus1pt \itemindent0pt
 \itemsep 4pt plus 2pt minus 1pt
 \advance\leftmargin\labelsep
 \usecounter{enumi}}
 \def\newblock{\hskip .11em plus .33em minus .07em}
 \sloppy\clubpenalty4000\widowpenalty4000
 \sfcode`\.=1000\relax}

\usepackage[centertags]{amsmath}
\usepackage{amsmath}

\usepackage[dutch,british]{babel}
\usepackage{amsfonts}
\usepackage{amssymb}
\usepackage{newlfont}

  \newtheorem{theorem}{Theorem}[section]
  \newtheorem{corollary}[theorem]{Corollary}

  \newtheorem{remark}[theorem]{Remark}
  \newtheorem{assumption}[theorem]{Assumption}

\numberwithin{equation}{section}

\DeclareMathOperator{\Tr}{Tr}

\newcommand\otimesal{\mathop{\hbox{\raise 1.6 ex
  \hbox{$\scriptscriptstyle\mathrm{al}$}
\kern -0.92 em \hbox{$\otimes$}}}}
\newcommand\oplusal{\mathop{\hbox{\raise 1.6 ex
  \hbox{$\scriptscriptstyle\mathrm{al}$}
\kern -0.92 em \hbox{$\oplus$}}}}
\newcommand\Gammal{\hbox{\raise 1.7 ex
\hbox{$\scriptscriptstyle\mathrm{al}$}\kern -0.50 em $\Gamma$}}
\renewcommand\i{\mathrm{i}}

 \let\Ga=\Gamma  
  
\newcommand\loplus{\mathop{\oplus}\limits}

\newcommand{\cD}{{\mathcal D}}

\newcommand{\cF}{{\mathcal F}}

\newcommand{\cH}{{\mathcal H}}

\newcommand{\cK}{{\mathcal K}}

\newcommand{\cN}{{\mathcal N}}

\newcommand{\cP}{{\mathcal P}}

\newcommand{\cZ}{{\mathcal Z}}
\renewcommand\c{{\mathrm c}}
\newcommand\cl{{\mathrm{cl}}}

\newcommand{\cc}{{\mathbb C}}

\newcommand{\rr}{{\mathbb R}}

\newcommand{\opunit}{\text{1}\kern-0.22em\text{l}}

\newcommand{\frh}{{\mathfrak h}}

\newcommand{\id}{{\mathrm{id}}}

\newcommand{\slim}{{\mathrm s-}\lim}
\newcommand{\e}{{\mathrm e}}

\renewcommand{\d}{{\mathrm d}}

\newcommand{\res}{{\mathrm R}}

\renewcommand{\sp}{\mathrm{sp}}
\newcommand{\Ran}{\mathrm{Ran}}

\newcommand{\Dom}{\mathrm{Dom}}

\newcommand{\beq}{ \begin{equation} }
\newcommand{\eeq}{ \end{equation} }
\newcommand{\bet}{ \begin{theorem} }
\newcommand{\eet}{ \end{theorem} }

\newcommand{\s}{{\mathrm s}}

 \newcounter{smallarabics}
\newenvironment{arabicenumerate}
{\begin{list}{{\normalfont\textrm{\arabic{smallarabics})}}}
  {\usecounter{smallarabics}\setlength{\itemindent}{0cm}
  \setlength{\leftmargin}{5ex}\setlength{\labelwidth}{4ex}
  \setlength{\topsep}{0.75\parsep}\setlength{\partopsep}{0ex}
   \setlength{\itemsep}{0ex}}}
{\end{list}}

\newcounter{smallroman}

\newcommand{\ben}{\begin{arabicenumerate}}
\newcommand{\een}{\end{arabicenumerate}}

\newcommand{\wdr}{}

\begin{document}



\abbrevauthors{J. Derezi\'{n}ski and W. De Roeck}

\abbrevtitle{Reduced and Extended  Weak Coupling Limit}

\title{Reduced and Extended \\ Weak Coupling Limit}

\author{Jan Derezi\'{n}ski}
\address{ Department of Mathematical Methods in Physics \\
Warsaw University \\
 Ho\.{z}a 74, 00-682, Warszawa, Poland\\
E-mail: jan.derezinski@fuw.edu.pl}

\author{Wojciech De Roeck}
\address{ Instituut voor Theoretische Fysica, K.U.Leuven\\
 Belgium \\
E-mail: wojciech.deroeck@fys.kuleuven.be}

\maketitlebcp

{\renewcommand\contentsname{}
\renewcommand{\settocname}{}

\tableofcontents
}

\section{Introduction.}

Physicists often describe quantum systems by
{\em completely positive} (c.p.)
  semigroups \cite{Haa,AL,Al2}.
It is generally believed that this approach
is only a phenomenological approximation to
a more fundamental description.
 One usually assumes that on the fundamental level
the dynamics of  quantum systems is unitary, more precisely, is
 of the form $t\mapsto \e^{\i tH}\cdot\e^{-\i
  tH}$ for some self-adjoint $H$.

One of justifications
for the use of c.p. semigroups in quantum physics
 is based on the so-called weak coupling limit for the reduced
dynamics  \cite{VH,Da1},
which we will call the {\em reduced weak coupling limit}.
One assumes that a {\em small system} is coupled to a {\em large reservoir} and the
dynamics of the full system is unitary. The
interaction between the small system and
the reservoir is multiplied by a small coupling
constant $\lambda$. Often the  reservoir is described by a {\em free Bose gas}.

The basic steps of the reduced weak coupling limit are:
\begin{itemize}\item
Reduce the dynamics to the small system.

\item Rescale time as $\frac{t}{\lambda^2}$.
\item Subtract the dynamics of the small system.
\item Consider the weak coupling $\lambda\to0$.

\end{itemize}
In the limit one obtains a dynamics given by a c.p. Markov
semigroup.

Another possible justification of c.p. dynamics goes as follows.
 One considers
 the tensor product of the small system and an appropriate bosonic
reservoir. On this enlarged space one constructs a certain unitary dynamics
 whose
 reduction to the small system is a c.p. semigroup.
We will call it a {\em  quantum Langevin  dynamics}. Another name used in
 this context in  the
 literature is a {\em  quantum  stochastic dynamics}. Its construction has a
 long history, let us mention \cite{AFLe,HP,Fr,Maa}.

Thus one can obtain a Markov  c.p. semigroup by reducing a
 {\em single} unitary dynamics,
 without invoking  a family of dynamics and taking its limit.
 However,
the generator of  a quantum Langevin dynamics equals $\i [Z,\cdot]$
where $Z$ is a self-adjoint operator that does not look like a
 physically realistic Hamiltonian. In particular, it
is unbounded from both below and
 above. Thus one can question the physical
 relevance of this construction.

It turns out, however, that one can extend the weak coupling limit in such a
way,  that it
involves not only the small system, but also the reservoir. As a result of
this approach one can obtain a quantum Langevin dynamics.
One can argue that
this approach gives a physical justification of quantum Langevin dynamics.

The above idea was  first
implemented in \cite{AFL}
 by Accardi, Frigerio and Lu under the name of the {\em stochastic
limit } (see also \cite{ALV}).
Recently we presented our version of this approach,
under the name of the
 {\em extended weak coupling limit} \cite{DD1,DD2}, which  we believe
 is simpler and more
natural that of \cite{AFL}.
The
basic steps of  the extended weak coupling limit are:
\begin{itemize}

\item Introduce the so-called  {\em
  asymptotic space}
  --- the tensor product of the space of the
  small system and of the {\em asymptotic reservoir}.
\item Introduce  an {\em identification operator}
 that maps the asymptotic reservoir
  into the physical reservoir  and
rescales its energy by   $\lambda^2$ around the
    Bohr frequencies.

    \item Rescale time as $\frac{t}{\lambda^2}$.

\item Subtract the ``fast degrees of freedom''.

    \item Consider the weak coupling $\lambda\to0$.

\end{itemize}
In the limit one obtains a  quantum Langevin dynamics on
the
  asymptotic space. Note that
the asymptotic reservoir is given by a bosonic Fock space (just as the
physical reservoir). Its states are however different --
correspond only to those physical bosons whose  energies  differ from the
 {\em Bohr frequencies} by at most $O(\lambda^2)$. Only such bosons
 survive the weak coupling limit.

Let us mention yet another  scheme of deriving quantum Langevin
equations that
has received  attention in the literature, namely the `repeated
interaction models' where the reservoir is continuously refreshed,
see \cite{AtP,AtJ}.


In this article we review various aspects of the weak coupling limit,
reduced and, especially, extended, mostly following our papers
\cite{DD1,DD2}. We also describe some background material, especially related
to completely positive semigroups, quantum Langevin dynamics and the Detailed
Balance Condition.

The plan of our article is as follows.
In Section \ref{Toy model of the  weak coupling limit}
  we  describe both
kinds of the weak coupling limit on a class of toy-examples -- the so-called
{\em Friedrichs Hamiltonians} and their dilations.
 They are less relevant physically  than the main model treated in our article
 -- the one based on  Pauli-Fierz operators. Nevertheless, they
illustrate some of the main ideas of this limit in a simple and
mathematically instructive context. This section is based on \cite{DD1}.

In Sections  \ref{Completely positive maps and semigroups}
we recall some facts about completely positive maps and
semigroups, sketching  proofs of the {\em Stinespring dilation} theorem \cite{St}
and of the
so-called {\em Lindblad form}
of the  generator of a c.p. semigroup \cite{Li,GKS}.
  In particular, we discuss the freedom
of choosing various terms in  the  Lindblad form. This question, which we have
not seen discussed in the literature,
 is relevant for the construction of  quantum Langevin dynamics  and
the  weak coupling limit.

C.p. semigroups that arise in the weak coupling limit have an
additional property -- they commute with the unitary dynamics
generated by the Hamiltonian $K$ of the small system -- for shortness we say
that they are {\em $K$-invariant}. If in addition
the reservoir is thermal, they satisfy another special
 property -- the so-called {\em Detailed Balance
Condition (DBC)}
 \cite{DF1,Ag,FGKV,Al1}.
 We devote a large part of Sect.
 \ref{Completely positive maps and semigroups} to an analysis of
 the $K$-invariance and the DBC. We show that the generator of a c.p. semigroup with these
 properties has some features that curiously resemble the {\em Tomita-Takesaki
 theory} and the {\em KMS condition}. Let us note that
in our article the DBC is considered jointly with
 the $K$-invariance, because c.p. semigroups obtained in the weak coupling
 limit always
 have the latter property.

In Section \ref{Bosonic reservoirs} we describe the terminology and notation
that we use to describe second-quantized bosonic reservoirs interacting with a
small system. In particular, we introduce {\em Pauli-Fierz operators}
\cite{DJ1} -- used
often (also under other names)
in the physics literature to describe physically realistic systems.
In Subsect.
\ref{Weak coupling limit for thermal reservoirs}
we discuss thermal reservoirs. In our definition of a thermal reservoir
at inverse temperature $\beta$ one needs to check a simple condition for the
interaction without explicitly invoking the concept of a KMS state on an
 operator algebra, or
of a thermal Araki-Woods representation of the CCR \cite{DF1,DJ1}. Of course, this
condition is closely related to the KMS property.

In
 Subsection \ref{Quantum Langevin dynamics}
we describe a construction of quantum Langevin dynamics.
We include a discussion of the so-called quadratic noises, even though they are
still not used in our version of the extended weak coupling limit. (See however
\cite{Go} for some partial results in the context of the formalism of
\cite{AFL}).

In Section \ref{Weak coupling limit for Pauli-Fierz operators}
we describe the two kinds of  the
weak coupling limit for Pauli-Fierz operators:  reduced and  extended.
Most of
this section is based on \cite{DD2}.

\section{Toy model of the  weak coupling limit.}
\label{Toy model of the  weak coupling limit}
This  section is
somewhat independent of the remaining part  of the article.
It
 explains the (reduced and extended)
 weak coupling limit in the
setting of contractive semigroups on a Hilbert space and their unitary
dilations. It gives us
an opportunity to explain some of the main ideas of the weak coupling
limit in a relatively simple setting. It is based on \cite{DD1}.

It is possible to construct physically
interesting
models based on the material of this section (e.g. by considering quadratic
Hamiltonians obtained by second quantization). We will  not discuss this
possibility further,
 since in the remaining part of the article we will
analyze more interesting and more realistic models.

\subsection{Dilations of contractive semigroups.}
\label{Dilations of contractive semigroups}

First let us recall the well known concept of a unitary dilation
of a contractive semigroup. Let $\cK$ be a Hilbert space and
$\e^{-\i t\Upsilon}$   a strongly continuous contractive
semigroup on $\cK$. This implies that $-\i \Upsilon$ is
dissipative: $-\i\Upsilon+\i\Upsilon^*\leq0$.

Let  $\cZ$  be a Hilbert space containing $\cK$, $I_\cK$ the embedding of
$\cK$ in $\cZ$ and $U_t$ a unitary group on $\cZ$. We say that
  $(\cZ,I_\cK,U_t)$ is a dilation of $\e^{-\i t\Upsilon}$ iff
\[I_\cK^*U_tI_\cK=\e^{-\i t\Upsilon},\ \ t\geq0.\]

It is well known that every weakly continuous contractive
semigroup possesses a unitary dilation (which is unique up to the
unitary equivalence
 if we additionally demand its minimality).
The original and well known construction of a unitary dilation is due to
 Foias and Nagy and can be found in \cite{NF} (see also \cite{EL}).
 Below we present another
construction, which looks   different from that of Foias-Nagy. Its
main idea is to view the generator of a unitary dilation as a kind
of
 a {\em singular
  Friedrichs operator}. (See the next section, where
 Friedrichs operators are introduced).
 Such a definition is well adapted to the extended
weak coupling limit.  The construction that we present
 seems to be less known in the mathematics literature
than that of Foias-Nagy. Nevertheless, similar constructions  are scattered in
the literature, especially in physics papers.

Let $\frh$ be an auxiliary space and $\nu\in B(\cK,\frh)$ satisfy
\beq \frac{1}{\i}(\Upsilon-\Upsilon^*)=-\nu^*\nu.\label{aux}\eeq
Note that such $\frh$ and $\nu$ always exist. One of  possible
choices is to take $\frh:=\cK$ and
$\nu:=\sqrt{\i(\Upsilon-\Upsilon^*)}$.

If $\phi$ is a vector, then $|\phi)$ will denote the operator
$\cc\ni\lambda\mapsto|\phi)\lambda :=\lambda \phi$.
Similarly, $(\phi|$
 will
denote its adjoint: $f\mapsto (\phi|f:=(\phi|f)\in\cc$.

 We will use a similar
notation also for unbounded functionals. For instance,
 $(1|$ will
 denote the (unbounded) linear functional on $L^2(\rr)$ given by
\beq (1|f=\int f(x)\d x\label{111}\eeq
 with the domain $L^2(\rr)\cap L^1(\rr)$.
$|1)$ will denote the  hermitian conjugate of $(1|$ in the sense of
sesquilinear forms: if $f\in L^2(\rr)\cap L^1(\rr)$, then
\[(f |1):=\int \bar f(x)\d x.\]

Let $Z_\res$ be the operator of multiplication on $L^2(\rr)$ by the variable
$x$.

Introduce the Hilbert spaces $\cZ_\res ;= \frh{\otimes}L^2(\rr)$ and $\cZ:=\cK\ \oplus\cZ_\res $.
Clearly, $\cK$ is contained in $\cZ$, so we have the obvious embedding
$I_\cK:\cK\to\cZ$. We also have the embedding
$I_\res:\cZ_\res\to\cZ$.

For $t\geq0$, consider the following sesquilinear form on $\cal K
\oplus (\frh \otimes (L^2(\rr)\cap L^1(\rr)))$:
\begin{eqnarray}
{U}_t &=& I_{\res} \e^{ -\i
t{Z}_{\res}}I_{\res}^*\ +\ I_\cK\e^{-\i t\Upsilon}I_\cK^* \label{kak}\\
&&- \i (2\pi)^{-\frac{1}{2}}I_\cK  \int^t_0 \, \d u \, \e^{-\i
(t-u)\Upsilon} \nu^*\otimes(1| \e^{-iu{Z}_{\res}}
I_{\res}^*\nonumber \\&&
 - \i(2\pi)^{-\frac{1}{2}}
I_{\res} \int^t_0 \, \d u \, \e^{-\i (t-u){Z}_{\res}}\nu\otimes|1)
 \e^{-\i u\Upsilon}I_\cK^*\nonumber\\
&&\!\!\!\!\!\!\!\!\!\!\!\!\!\!\!\!\!\!\!\!\!\!\!\!\!- (2\pi)^{-1} I_{\res}
 \!\!\!\!\!\!\!\!\!\!\!\!\!\!\!\! \int\limits_{0\leq u_1,u_2,  \,
   u_1+u_2\leq t}
\!\!\!\!\!\!\!\!\!\!\!\!\!\!\!
 \d u_1 \d u_2 \, \e^{-
\i u_2{Z}_{\res}}\nu\otimes|1)\e^{-\i
(t-u_2-u_1)\Upsilon}\nu^*\otimes(1|
 \e^{- \i u_1
{Z}_{\res}}I_{\res}^*.\nonumber
\end{eqnarray}

By a straightforward computation we obtain \cite{DD1} \bet
The form $U_t$ extends to a strongly continuous unitary
group and \[I_\cK^* U_t I_\cK=\e^{-\i t\Upsilon}, \ \ t\geq0.\]
 Thus
$(\cZ,I_\cK,U_t)$ is a dilation of $\e^{-\i t\Upsilon}$. \eet

Let $-\i Z$ denote the generator of $U_t$, so that $U_t=\e^{-\i
tZ}$.  $Z$ is a self-adjoint operator with a number of interesting
properties. It is not easy to describe it with a well-defined
formula. Formally it is given by the sesquilinear form \beq \left
[ \begin{array}{cc} \frac12(\Upsilon+\Upsilon^*) &
    (2\pi)^{-\frac{1}{2}}
\nu^*\otimes(1|
\\
(2\pi)^{-\frac{1}{2}} \nu\otimes|1) & Z_\res \end{array}
  \right].\label{expo}\eeq
Note that (\ref{expo}) looks like a special case of a Friedrichs
operator (see
Subsection \ref{Weak coupling limit for Friedrichs operators}
and \cite{DF2}). As it stands, however,  (\ref{expo})
does not define a unique self-adjoint operator. Nevertheless, we will
sometimes use the expression  (\ref{expo})  when referring to $Z$.

Note that it is possible to give a compact formula for the resolvent of $Z$,
(which is another possible method of defining $Z$).
For $z\in\cc_+$,
\begin{eqnarray*}
\label{definitionresolvent}
(z-Z)^{-1}&
:=&I_\res(z-{Z}_{\res})^{-1 }I_\res^*+
I_\cK (z-\Upsilon)^{-1}I_\cK^*\\
&&+(2\pi)^{-\frac{1}{2}} I_\cK (z-\Upsilon)^{-1}
\nu^*\otimes(1|(z-Z_{\res})^{-1}I_\res^*
\\&&+
(2\pi)^{-\frac{1}{2}} I_\res (z-Z_{\res})^{-1}\nu\otimes|1)(z-\Upsilon)^{-1}I_\cK^*\\
&&+(2\pi)^{-1} I_\res\  (z-Z_{\res})^{-1}\nu\otimes|1)\
(z-\Upsilon)^{-1}\ \nu^*\otimes(1| (z-Z_{\res})^{-1}\ I_\res^*.
\end{eqnarray*}

Yet another approach that allows to define $Z$ involves a ``cut-off procedure''.
In fact, $Z$ is the norm resolvent limit for $r\to\infty$
 of the following regularized operators:
\[Z_r:=\left [ \begin{array}{cc} \frac12(\Upsilon+\Upsilon^*) &
  (2\pi)^{-\frac{1}{2}} \nu^*\otimes(1|1_{[-r,r]}(Z_\res)
\\
(2\pi)^{-\frac{1}{2}} \nu\otimes 1_{[-r,r]}(Z_\res)|1) & 1_{[-r,r]}(Z_\res)Z_\res \end{array}
  \right].\]
Note that it is important to remove the  cut-off in a  symmetric way. If we
replace $[-r,r]$ with $[-r, ar]$    we usually obtain a different operator.
The convergence of $Z_r$ to $Z$ is the reason why we can treat
(\ref{expo}) as the formal expression for $Z$.

Next, let us mention a certain  invariance property of $Z$.
For $\lambda\in\rr$, introduce the following unitary operator on
$\cZ$
\[j_\lambda u=u,\ \ u\in\cK;\ \ \ \ \ j_\lambda
g(y):=\lambda^{-1}g(\lambda^{-2}y),\ \ g\in \cZ_{\res}.\] Note
that \[j_\lambda^*Z_{\res}j_\lambda=\lambda^2 Z_{\res},\ \ \
j_\lambda^*|1)=\lambda|1).\] Therefore, the operator $Z$ is
invariant with respect to the following scaling: \beq
Z=\lambda^{-2}  j_\lambda^* \left[\begin{array}{cc}
\lambda^2\frac12(\Upsilon+\Upsilon^*)&
\lambda(2\pi)^{-\frac{1}{2}} \nu^*\otimes(1|
\\
 \lambda (2\pi)^{-\frac{1}{2}}\nu\otimes|1) & {Z}_{\res} \end{array}
  \right]j_\lambda
.\label{inva}\eeq
(\ref{inva})
 will play an important role in the extended weak coupling limit.

Note that in the weak coupling limit it is convenient to use the
representation of
$Z_\res$ as a multiplication operator. Another  natural possibility is to
represent it as the differentiation operator.
Let us describe this alternative version of the dilation.

 The
 (unitary) Fourier
transformation on $\frh{\otimes}L^2(\rr)$ will be denoted as follows:
 \beq
\cF f(\tau):=(2\pi)^{-1/2}\int f(x)\e^{-\i \tau x}\d x.\label{fou}\eeq
We will use $\tau$ as the generic variable after the application of $\cF$.
 The operator $Z$
 transformed by  $1_\cK\oplus\cF$  will be denoted
\begin{eqnarray}
\hat Z&:= &(1_\cK\oplus\cF) Z(1_\cK\oplus\cF^*).
\end{eqnarray}
Introduce
 \beq D_\tau:= \frac{1}{\i}\partial_\tau.\label{defo1}\eeq
 Let
$(\delta_0|$ have the meaning of an (unbounded) linear
 functional on $L^2(\rr)$ with the domain, say, the first
Sobolev space $H^1(\rr)$,  such that \beq (\delta_0 |
f):=f(0).\label{defo2}\eeq
 Similarly, $|\delta_0)$ let be its hermitian adjoint in the sense of forms.
By applying the Fourier transform to (\ref{expo}), we can write
\begin{eqnarray}
\hat Z &= &\left [ \begin{array}{cc} \frac12(\Upsilon+\Upsilon^*)
&
    \nu^*\otimes(\delta_0 |
\\[2mm]
 \nu\otimes|\delta_0) & D_\tau \end{array}
  \right].\label{expoa}\end{eqnarray}
Clearly, $\e^{-\i t\hat Z}$ is also a dilation  of $\e^{-\i
t\Upsilon}$.

The operator $\hat Z$ (or $Z$)
 and the unitary group it generates has a number of curious
and confusing
properties. Let us describe one of them.
Consider the space
 $\cD:=\cK\oplus(\frh{\otimes}H^1(\rr))$. Clearly, it is a dense
subspace of $\cZ$. Let us define the following   quadratic form
on $\cD$: \beq \hat Z^+:=\left [ \begin{array}{cc} \Upsilon &
\nu^*\otimes(\delta_0|
\\
 \nu\otimes|\delta_0) & D_\tau \end{array}
  \right].\label{noise}\eeq
 Then, for $\psi,\psi' \in \cD$, \beq \lim_{t \downarrow
0}\frac{1}{t}
 ( \psi| (\e^{-\i t\hat Z}-1)  \psi' )=
-\i( \psi| {\hat Z}^+ \psi' ). \label{kah}\eeq
 One could think that $\hat Z^+=\hat Z$. But
$\hat Z^+$ is in general non-self-adjoint, which is incompatible
with the
 fact that $\e^{-\i t\hat Z} $ is a unitary group.

To explain this paradox we notice that
 $(\psi  |\e^{-\i t\hat Z} \psi')$ is in general
 not differentiable at zero: its right and left derivatives exist but are
 different. Hence
$\cD$ is not contained in the domain of the generator of $\hat Z$.
We will
 call $\hat Z^+$ the {\em false  form of the generator of} $\e^{\i
 t\hat Z}$.

In order to make an even closer contact with the usual form of the
quantum Langevin equation \cite{HP,At,Fa,Bar,Me}, define the
cocycle unitary \beq \hat W(t) := \e^{\i t D_\tau} \e^{-\i t\hat Z}.
\eeq

Then for   $t>0$, or  for $t=0$ and the right derivative, we have
the ``toy Langevin (stochastic) equation'' which  holds in the
sense of quadratic forms on $\cal D$, \beq \i\frac{\d}{\d t} \hat W(t)
=
 ({\Upsilon}+ \nu \otimes |\delta_t ))\hat  W(t)
 +  \nu^*{\otimes} (\delta_t| .
\label{kah2}\eeq
%

\subsection{``Toy quadratic noises''.}
\label{``Toy quadratic noises''}

The formula for $\hat Z$ or for $\hat Z^+$ has one interesting feature: it
involves a perturbation that is localized just at  $\tau=0$.
 One can ask whether one can consider other dilations with more
general perturbations localized at $\tau=0$. In this subsection we will
describe such dilations. This construction will not be  needed in the present
version of the weak
coupling limit. We believe it is an interesting
``toy version'' of ``quadratic noises'', which we will discuss
in Subsect \ref{Quadratic noises}.
We also expect to extend the results of \cite{DD1} to ``toy quadratic noises''.

Clearly, for any unitary operator $U$ on $\frh{\otimes}L^2(\rr)$,
$(1_\cK\oplus U)\e^{\i t Z}(1_\cK\oplus U^*)$ is a dilation of $\e^{-\i
  t\Upsilon}$. Let us choose a special $U$, which will lead to a perturbation
localized at $\tau=0$.

Let $S$ be a unitary operator on
$\frh$. For $\psi\in\frh{\otimes}L^2(\rr)\simeq L^2(\rr,\frh)$ we set
\beq \gamma(S) \psi(\tau):=\left\{\begin{array}{ll}S\psi(\tau),&\tau>0,\\
\psi(\tau),&\tau\leq0.
\end{array}\right.\label{qua1}\eeq
Then $\gamma(S)$ is a unitary operator on $\frh{\otimes}L^2(\rr)$.
Set
\[\hat Z_S:=(1_\cK\oplus \gamma(S)^*) \hat Z(1_\cK\oplus \gamma(S)).\]
Clearly, $\e^{\i t\hat{Z}_S }$  is a dilation of $\e^{-\i
t\Upsilon}$.
 It is awkward to
write down a formula for $\hat{Z}_S$
in the matrix form, even just formally. It is more natural
 to write
down the ``false form of $Z_S$'':
\begin{eqnarray*}
\hat Z_S^+&:=&
(1_\cK\oplus \gamma(S)^*) \hat Z^+(1_\cK\oplus \gamma(S))\\[2ex]
&=&\left [ \begin{array}{cc} \Upsilon &
    \nu^*S\otimes(\delta_0 |
\\
 \nu\otimes|\delta_0) & D_\tau+\i(1-S){\otimes}|\delta_0)(\delta_0|
 \end{array}\right].\end{eqnarray*}
For $\psi,\psi'\in\cD$  we have \beq
\lim_{t\downarrow 0}\frac{1}{t}
 ( \psi| (\e^{-\i t\hat Z_S} -1) \psi' )=
-\i( \psi| \hat{Z}_S^+ \psi' ), \label{qua}\eeq Again, as in
(\ref{kah2}), one can extend this formula to derivatives at $t>0$.
Let \beq \hat W_S(t):= \e^{\i tD_\tau}  \e^{-\i t\hat Z_S}, \eeq then,
 in the sense of quadratic forms on $\cal D$, \beq
\i\frac{\d}{\d t} \hat W_S(t) =
 ({\Upsilon}+ \nu \otimes |\delta_t ))\hat W_S(t)
 +  \nu^* S {\otimes}(\delta_t| +\i(1-S){\otimes}|\delta_t)(\delta_t| .
\eeq

\subsection{Weak coupling limit for Friedrichs operators.}
\label{Weak coupling limit for Friedrichs operators}

Let $\cH:=\cK\oplus\cH_\res$ be a Hilbert space, where $\cK$ is finite
dimensional.  Let $I_\cK$ be the embedding
of $\cK$ in $\cH$. Let $K$ be a self-adjoint operator on $\cK$ and $H_\res$ be
a self-adjoint operator on $\cH_\res$. Let $V$ be a linear operator from
$\cK$ to $H_\res$.
The following class of operators will be called {\em Friedrichs operators}:
\[H_\lambda
:
=\left [ \begin{array}{cc} K &
    \lambda V^*
\\    \lambda V&
  H_\res \end{array}
  \right].\]

Assume that $\int\|V^*\e^{-\i tH_\res}V\|\d t<\infty$. Then we can
define the following operator, sometimes called the {\em Level
Shift Operator}, since it describes the shift of eigenvalues of
$H_\lambda$ in perturbation theory at the 2nd order in $\lambda$:
\beq \label{def: abstract Gamma}\Upsilon:=\sum_{k\in\sp
K}\int_0^\infty 1_k(K)V^*\e^{-\i t(H_\res-k)}V1_k(K)\d t, \eeq
where $1_k(K)$ denotes the spectral projection of $K$ onto the
eigenvalue $k$; $\sp K$ denotes the spectrum of $K$. Note that
$\Upsilon K=K\Upsilon$.

The following theorem is a special case of a result of Davies
\cite{Da1,Da2,Da3}, see also \cite{DD1}:
\bet[Reduced weak coupling limit for Friedrichs operators]
\[\lim_{\lambda\to0}\e^{\i tK/\lambda^2}I_\cK^*\e^{-\i tH_\lambda/\lambda^2}I_\cK
=\e^{-\i t\Upsilon}.\] \eet

In order to study the extended weak coupling limit for Friedrichs operators we
need to make additional assumptions. They are perhaps a little complicated to
state, but they are satisfied in many concrete situations.
\begin{assumption}
We suppose that for any $k\in\sp K$ there exists an open
$I_k\subset\rr$ and a Hilbert space $\frh_k$ such that $k\in I_k$,
\begin{eqnarray*}
\Ran 1_{I_k}(H_\res)&\simeq &\frh_k\otimes L^2(I_k,\d x),
\end{eqnarray*}
$1_{I_k}(H_\res)H_\res $ is the multiplication operator by the variable
 $x\in I_k$ and
\begin{eqnarray*}
 1_{I_k}(H_\res) V&\simeq & \int_{I_k}^\oplus v(x)\d x.
\end{eqnarray*}
We assume that $I_k$ are disjoint for distinct $k$ and the measurable function
$I_k \ni x\mapsto v(x)\in B(\cK,\frh_k)$ is
 continuous at $k$.
\end{assumption}

In other words, we assume  that the reservoir Hamiltonian $H_\res$ and
the interaction $V$ are ``nice'' around the spectrum of $K$. In fact, in the
extended weak coupling limit only a vicinity of $\sp K$ matters.

We set $\frh:=\loplus_k\frh_k$, $\cZ_\res := \frh{\otimes} L^2(\rr)$ and
$\cZ:=\cK\oplus \cZ_\res.$ $\cZ_\res$ and
$\cZ$ are   the so-called {\em
  asymptotic
 spaces},
 which are in general different from the {\em physical spaces} $\cH_\res$ and
$\mathcal H$.

Next, let us describe the
{\em asymptotic
  dynamics}.
Let $\nu:\cK\to\frh$ be defined as
\[\nu:=(2\pi)^{\frac{1}{2}}\loplus_k v(k)1_k(K).\]
Note that it satisfies (\ref{aux}) with $\Upsilon$ defined by
(\ref{def: abstract Gamma}). This follows by extending the
integration in  (\ref{def: abstract Gamma}) to $\rr$ and using the
inverse Fourier transform.   As before, we set $Z_\res$ to be the
multiplication by $x$ on $L^2(\rr)$ and we define $\e^{-\i tZ}$ by
(\ref{kak}), so that $(\cZ,I_\cK,\e^{-\i tZ})$ is a dilation of
$\e^{-\i t\Upsilon}$.

Finally, we need an
 identification operator that maps  the asymptotic space into the
physical space. This is the least canonical part of the construction. In fact,
 there is some arbitrariness in its definition for the frequencies away from
 $\sp K$.
For $\lambda > 0$, we define the family of partial isometries
$J_{\lambda,k}:  L^2(\rr, \frh_k)
 \rightarrow   L^2(I_k, \frh_k)\subset \cH$:
\begin{equation*}
  (J_{\lambda,k}g_k)(y)=   \left\{ \begin{array}{ll}
        \frac{1}{\lambda}
        g_k(\frac{y-k}{\lambda^2}),  & \textrm{ if }   y\in
        I_k;
 \\
        0,  & \textrm{ if } y\in\rr\backslash I_k. \\ \end{array}
  \right.
\end{equation*}
We set $J_\lambda: \cZ\to\cH$, defined for $g=(g_k)\in\cZ_\res$ by
\[J_\lambda g:=\sum_k J_{\lambda,k}g_k,\]
and on $\cK$ equal to the identity.
Note that $J_\lambda$ are partial isometries and
\[\slim_{\lambda\searrow0}J_\lambda^*J_\lambda=1.\]


The following result is proven in \cite{DD1}:
\bet[Extended weak coupling limit for Friedrichs operators]
\begin{eqnarray*}
&&  \s^*-\lim_{\lambda \searrow 0}
 J^*_{\lambda}  \e^{\i\lambda^{-2} t H_0}
\e^{\lambda^{-2}(t-t_0) H_\lambda } \e^{\i\lambda^{-2} t_0  H_0}
 J_\lambda  =\e^{\i t Z_\res}\e^{-\i (t-t_0)Z }\e^{\i t_0 Z_\res}.
\end{eqnarray*}
\eet

Here we used the  strong* limit: $\s^*-\lim_{\lambda \searrow 0}
A_\lambda=A$ means that for any vector $\psi$ we have
$\lim_{\lambda \searrow 0} A_\lambda\psi=A\psi$,
$\lim_{\lambda \searrow 0} A_\lambda^*\psi=A^*\psi$.

Note that in the extended weak coupling limit for Friedrichs operators
the asymptotic space is a
direct sum of parts belonging to
 various
eigenvalues of $K$ that ``do not talk to one another''--have independent
asymptotic dynamics.

\section{Completely positive maps and semigroups.}
\label{Completely positive maps and semigroups}

This section  presents basic material about completely positive
maps and semigroups.
In particular, we describe a construction of
 the Stinespring dilation \cite{St}
 and of the
so-called Lindblad form
of the  generator of a c.p. semigroup \cite{Li,GKS}.
These beautiful classic results are  described in many places in the
literature. Nevertheless, some  of their aspects,
mostly
concerning
the freedom of choice of various terms in the Lindblad form, are difficult to
find in the literature. Therefore, we
describe this material at length, including sketches of proofs.

In Subsect. \ref{Classical Markov semigroups} we recall the usual
concept  of a (classical)
Markov semigroups (on a finite state space). When discussing c.p. (quantum)
Markov semigroups, it is useful to compare it to their  classical
analogs, which are usually much simpler.

In Subsect \ref{Invariant c.p semigroups}
we discuss c.p. semigroups invariant
with respect to a certain unitary dynamics. Such c.p.
semigroups arise in the weak
coupling limit -- therefore, one can argue that they are ``more physical than
others''.

 Finally, in Subsect.
\ref{Detailed Balance Condition} we analyze
the Detailed Balance
Condition, which singles out c.p. dynamics obtained from a thermal
reservoir.

\subsection{Completely positive maps.}
\label{Completely positive maps}

Let $\cK_1,\cK_2$ be Hilbert spaces. We say that a map
$\Xi: B(\cK_1)\to B(\cK_2)$
is positive iff $A\geq0$ implies $\Xi(A)\geq0$.
We say that $\Xi$ is Markov iff $\Xi(1)=1$.
We say that a map $\Xi$ is $n$-positive
iff
\[\Xi\otimes \id: B(\cK_1\otimes\cc^n)\to B(\cK_2\otimes\cc^n)\]
is positive. ($\id$ denotes the identity).
We say that $\Xi$ is  completely positive, or  c.p. for
short, iff it is $n$-positive for any $n$.

It is easy to see that
 if $\frh$ be a Hilbert space and $\nu\in
B(\cK_2,\cK_1\otimes\frh)$. Then
\beq\Xi(A):=\nu^*\  A{\otimes} 1\ \nu \label{*}\eeq
is c.p. The following theorem says that the above representation of a c.p. map
 is universal.
2) means that this representation is unique up to a unitary
isomorphism.


\bet[Stinespring] Assume that  $\cK_1,\cK_2$ are finite
dimensional.
 \ben
\item
If $\Xi$ is c.p.
from $B(\cK_1)$ to $B(\cK_2)$,  then there exist
a Hilbert space $\frh$ and $\nu
\in
B(\cK_2,\cK_1\otimes\frh)$  such that (\ref{*}) is true and
\beq\{(\phi|{\otimes}1_\frh\ \nu\ \psi\ :\ \phi\in\cK_1,\
\psi\in\cK_2\}=\frh.\label{minio}\eeq
\item
If in addition to the $\frh'$ and $\nu'$ also satisfy the
above properties, then there exists a
unique unitary operator $U$
from $\frh$ to $\frh'$ such that $\nu'=1_{\cK_1}\otimes U\ \nu$.
\een\label{stinespring}\eet

The right hand side of
 (\ref{*})  is called a {\em Stinespring dilation} of a c.p. map $\Xi$.
If the condition (\ref{minio}) holds, then it is called a {\em minimal}.

\begin{remark}
If we choose
 a basis in $\frh$, so that we identify $\frh$ with $\cc^n$, then we can
 identify $\nu$ with $\nu_1,\dots,\nu_n\in B(\cK_2,\cK_1)$. Then we can
 rewrite  (\ref{*}) as
\beq \Xi(A)=\sum_{j=1}^n\nu_j^*A\nu_j.\label{kraus}\eeq
In the literature, (\ref{kraus}) is called a {\em Kraus decomposition}, even
 though the work of Stinespring is much earlier than that of Kraus.
\end{remark}

Note that physically
the space $\frh$ can be interpreted as a part of the reservoir that
directly interacts with the small system.

\noindent{\em Proof of Theorem \ref{stinespring}.} Let us prove 1).
We equip the algebraic tensor product $\cH_0:=B(\cK_1)\otimes\cK_2$ with the
following scalar
product: for
\[\tilde v=\sum_iX_i\otimes v_i,\ \ \ \tilde w=\sum_iY_i\otimes w_i\in\cH_0
\] we set
\begin{eqnarray*}
(\tilde v|\tilde w)&
=&\sum_{i,j}(v_i|\Xi(X_i^*Y_j)w_j).\end{eqnarray*}
By the complete positivity, it is positive definite.

Next we note that there exists a unique linear map
$\pi_0:B(\cK_1)\to B(\cH_0)$ satisfying
\[\pi_0(A)\tilde v:=\sum_i AX_i\otimes v_i.\]
We check that
\begin{eqnarray*}
(\pi_0(A)\tilde v|\pi_0(A)\tilde v)\leq \|A\|^2(\tilde v|\tilde v),&
\pi_0(AB)=\pi_0(A)\pi_0(B),&
\pi_0(A^*)=\pi_0(A)^*.
\end{eqnarray*}

Let $\cN$ be the set of $\tilde v\in\cH_0$ with $(\tilde v|\tilde v)=0$.
Then the completion of  $\cH:=\cH_0/\cN$ is a Hilbert space.
There exists a nondegenerate
 $*$-representation $\pi$ of $B(\cK_1)$ in \wdr{$B(\cH)$} such that
\[\pi(A)(\tilde v+\cN)=\pi_0(A)\tilde v.\]
Using the fact that all our spaces are finite dimensional we see
that  for some Hilbert space $\frh$
we can identify $\cH$ with $ \cK_1\otimes\frh$
and $\pi(A)=A\otimes 1$.

We set
\[\nu v:=1\otimes v+\cN.\]
We check that
\[\Xi(A)=\nu^* A{\otimes} 1\ \nu.\]
This ends the proof of the existence of the Stinespring dilation.

Let us now prove 2).
If $\frh'$, $\nu'$ is another pair that gives a Stinespring dilation,
we check that
\[\left\|\sum_i X_i\otimes1_\frh\ \nu\ v_i\right\|
=\left\|\sum_i X_i\otimes1_{\frh'}\ \nu'\ v_i\right\|.\]
Therefore, there exists a unitary $U_0: \cK_2\otimes\frh\to\cK_2\otimes\frh'$
such that $U_0\nu=\nu'$. We check that
$U_0\ A\otimes1_\frh= A\otimes 1_{\frh'}\  U_0$. Therefore, there exists a
unitary
$U:\frh\to\frh' $
such that $U_0=1\otimes U$. \endproof

We will need the following inequality for c.p. maps:

\bet[Kadison-Schwarz inequality for c.p. maps.]
 If $\Xi$ is 2-positive \\ and $\Xi(1)$ is invertible, then
\beq\Xi(A)^*\Xi(1)^{-1}\Xi(A)\leq\Xi(A^*A).\label{kadi}\eeq
\eet

\Proof
Let $z\in\cc$. $\left[\begin{array}{cc}A^*A&\ zA^*\\
    \bar z A&|z|^2\end{array}\right] \geq0$ implies
 $\left[\begin{array}{cc}\Xi(A^*A)& z\Xi(A^*)\\
    \bar z\Xi(A)&|z|^2\Xi(1)\end{array}\right] \geq0$. Hence, for
    $\phi,\psi\in\cK$,
\beq (\phi|\Xi(A^*A)\phi)+2{\mathrm{Re}}\wdr{ \bar z
(\psi|\Xi(1)^{-1/2}\Xi(A)\phi)}+|z|^2(\psi|\psi)\geq0.\eeq
Therefore, \beq
(\phi|\Xi(A^*A)|\phi)
(\psi|\psi)\geq|(\psi|\Xi(1)^{-1/2}\Xi(A)\phi)|^2,\eeq
which implies (\ref{kadi}). \endproof


\subsection{Completely positive semigroups.}
\label{Completely positive semigroups}

Let $\cK$ be a finite dimensional Hilbert space.
Let us  consider a c.p. semigroup  on $B(\cK)$.
We will always assume the semigroup to be continuous, so that it can be
written as $\e^{tM}$ for a bounded operator $M$ on $B(\cK)$.
We will  call  $\e^{tM}$
 Markov  if it preserves the identity.

C.p. Markov semigroups appear in the literature under various
names. Among them let us mention {\em quantum Markov semigroups} and
{\em quantum dynamical semigroups}.

If $M_1$, $M_2$ are the generators of (Markov)
 c.p. semigroups and $c_1,c_2\geq0$, then
$c_1M_1+c_2M_2$ is the  generator of a (Markov) c.p. semigroup.
This follows by the Trotter formula.

Here are two classes of
examples of c.p. semigroups:
\ben\item
 Let $\Upsilon=\Theta+\i\Delta$ be an operator on $\cK$, with $\Theta,\Delta$
 self-adjoint.  Then
\[M(A):=\i \Upsilon A-\i A\Upsilon^*=\i[\Theta,A]-[\Delta,A]_+\] is the generator of
a c.p.  semigroup and
\[\e^{t M}(A)=\e^{\i t\Upsilon} A\e^{-\i t \Upsilon^*}.\]
\item
Let $\Xi$ be a c.p. map on $B(\cK)$. Then it is the generator of a c.p. semigroup
and
\[\e^{t\Xi}(A)=\sum_{j=0}^\infty\frac{t^j}{j!}\Xi^j(A).\]
\een

Let $\Theta$, $\Delta$ be self-adjoint operators on $\cK$.
Let $\frh$ be  an auxiliary Hilbert space and
 $\nu\in B(\cK,\cK\otimes\frh)$.
Then it follows from what we wrote above that
\beq M(S)= \i[\Theta,A] -[\Delta, A]_++  \nu^*\ A{\otimes}1\ \nu,   \qquad A \in
B(\cK),\label{lindi}\eeq
is the generator of a c.p. semigroup.
$\e^{tM}$ is Markov iff $2\Delta=\nu^*\nu$.

The following theorem gives a complete characterization of generators of
c.p. semigroups on a finite dimensional space \cite{Li,GKS}.

\bet[Lindblad, Gorini-Kossakowski-Sudarshan]
\ben\item
 Let $\e^{tM}$ be a\\ c.p. semigroup on
 $B(\cK)$ for   a finite dimensional Hilbert space $\cK$.
 Then
there exist self-adjoint operators $\Theta$, $\Delta$ on $\cK$,
 an auxiliary Hilbert space $\frh$ and an
 operator $\nu\in B(\cK,\cK\otimes\frh)$
such that $M$ can be written
 in the form (\ref{lindi}) and
\beq
\{(\phi|{\otimes}1\ \nu\ \psi\ :\ \phi,\psi\in\cK\}=\frh
.\label{miniu}\eeq
\item
We can  always
choose $\Theta$ and $\nu$ so that
\[\Tr\Theta =0,\ \ \ \Tr\nu=0.\]
(Above, we take the trace of $\nu$ on the space $\cK$ obtaining a vector in
$\frh$).
If this is the case, then $\Theta$ and $\Delta$ are determined uniquely, and
$\nu$ is determined uniquely up to the unitary equivalence.\een
\label{lindo}\eet

We will say that a c.p. semigroup is {\em purely
  dissipative} if
$\Theta=0$. We will call (\ref{lindi}) a {\em Lindblad form } of $M$. We will
  say that it is {\em minimal} iff (\ref{miniu}) holds.

\begin{remark} If we identify $\frh$ with $\cc^n$, then we can write
\[\nu^*\ A{\otimes}1\ \nu=\sum_{j=1}^n\nu_j^*A\nu_j.\]
Then $\Tr\nu=0$ means $\Tr\nu_j=0$, $j=1,\dots,n$.
\end{remark}

\noindent{\em Proof of Theorem \ref{lindo}.}
Let us prove 1).
The unitary group on $\cK$, denoted $U(\cK)$, is compact. Therefore,
there exists the Haar measure on $U(\cK)$, which
we  denote $\d U$.
Note that
\[\int UXU^*\d U=\Tr X.\]

Define
\[\i\Theta- \Delta_0:=\int M(U^*)U\d U,\]
where $\Theta$ and $\Delta_0$ are self-adjoint.

Let us show that
\beq \int M(XU^*)U\d U=(\i\Theta-\Delta_0) X.\label{nmn}\eeq
 First check this identity for unitary $X$, which follows by
the invariance of the measure $\d U$. But
every operator is a linear combination of unitaries. So (\ref{nmn}) follows in
general.

We can apply the Kadison-Schwarz inequality  to the semigroup $\e^{tM}$:
\beq \e^{tM}(X)^*\e^{tM}(1)^{-1}\e^{tM}(X)\leq\e^{tM}(X^*X).\label{schw3}\eeq
Differentiating  (\ref{schw3}) at $t=0$ yields
\beq M(X^*X)+X^*M(1)X-M(X^*)X-X^*M(X)\geq0.\label{schw}\eeq
Replacing $X$ with $UX$, where $U$ is unitary, we obtain
\beq M(X^*X)+X^*U^*M(1)UX-M(X^*U^*)UX-X^*U^*M(UX)\geq0.\label{schw1}\eeq
Integrating (\ref{schw1})  over $U(\cK)$ we get
\beq
M(X^*X)+X^*X\Tr
 M(1)-(\i\Theta-\Delta_0)X^*X-X^*X(-\i\Theta-\Delta_0)^*\geq0.\label{schw2}\eeq

Define
\begin{eqnarray*}\Delta_1&:=&\Delta_0+\frac12\Tr M(1),\\
\Xi(A)&:=&M(A)-(\i\Theta-\Delta_1)A-A(-\i\Theta-\Delta_1).
\end{eqnarray*}
Using (\ref{schw2}) we see that  $\Xi$ is positive. A straightforward
 extension of the above argument shows that $\Xi$ is also
 completely positive. Hence, by Theorem \ref{stinespring} 1),
 it can be written as
\[\Xi(A)=\nu_1^* \ A{\otimes}1\ \nu_1,\]
for some auxiliary  Hilbert space $\frh$ and a map
$\nu_1:\cK\to\cK{\otimes}\frh$.

Finally, let us prove 2).
The operator $\Theta$ has trace zero, because
\begin{eqnarray*}
\i\Tr \Theta-\Tr \Delta_0&=&
\int U_1M(U^*)UU_1^*\d U\d U_1\\
&=&
\int U_2UM(U^*)U_2^*\d U\d U_2\\
&=&-\i\Tr \Theta-\Tr \Delta_0.
\end{eqnarray*}

Let $w$ be an arbitrary vector in $\frh$ and
\begin{eqnarray*}
\Delta&:=&\Delta_1+\nu^* 1{\otimes}|w)+\frac12(w|w),\\
\nu&:=&
\nu_1+1 {\otimes}|w).\end{eqnarray*}

Then the same generator of a c.p. semigroup can be written in two Lindblad forms:
\begin{eqnarray*}
&&(\i\Theta-\Delta_1)A+A(-\i\Theta-\Delta_1)+\nu_1^*A\nu_1,\\
&=&
(\i\Theta-\Delta)A+A(-\i\Theta-\Delta)+\nu^*A\nu.
\end{eqnarray*}

In particular, choosing
 $w:=-\Tr \nu_1$,
we can make sure that $\Tr\nu=0$.
\endproof

\subsection{Classical Markov semigroups.}
\label{Classical Markov semigroups}

It is instructive to compare c.p. Markov semigroups with usual (classical)
Markov semigroups.

Consider the space $\cc^n$. For $u=(u_1,\dots,u_n)\in \cc^n$ we
will write $u\geq0$ iff $u_1,\dots,u_n\geq0$. We define ${\mathbf
1}:=(1,\dots,1)$. We say that a linear map $T$ is {\em pointwise
positive } iff $u\geq0$ implies $Tu\geq0$. We say that it is {\em
Markov}  iff $T{\mathbf 1}={\mathbf 1}$.

A one-parameter semigroup
 $\rr_t\mapsto T_t\in B(\cc^n)$ will be  called a {\em (classical) Markov
  semigroup} iff  $T_t$
is pointwise positive and Markov for any $t\geq0$.

Every continuous one-parameter semigroup on $\cc^n$ is of the form $\rr_+\ni
t\mapsto\e^{tm}$
for some
  $n\times n$ matrix $m$.
Clearly, the transformations
$\e^{tm}$ are pointwise
positive for any $t\geq0$
iff $m_{ij}\geq0$, $i\neq j$. They are  Markov  for any $t\geq0$ iff in
addition $\sum_j m_{ij}=0$.

Markov c.p.
 semigroups often lead to classical Markov semigroups, as described in the
following easy fact:
\bet Let $P_1,\dots,P_n\in B(\cK)$ satisfy $P_j^*=P_j$ and
$P_jP_k=\delta_{jk}P_j$. Let $\cP$ be the (commutative) $*$-algebra generated
by $P_1,\dots,P_n$. Clearly, $\cP$ is naturally isomorphic to $\cc^n$.
Let $\e^{tM}$ be a Markov c.p. semigroup on $B(\cK)$ that preserves the algebra
$\cP$. Then $\e^{tM}\Big|_\cP$ is a classical Markov
semigroup.\label{dbc1} \eet

Conversely, from a classical Markov semigroup one can construct
c.p. Markov semigroups:

\bet Let $\e^{tm}$ be a
 classical Markov semigroup
on $\cc^n$.
Let
$e_1,\dots,e_n$ denote the canonical basis of $\cc^n$ and
$E_{ij}:=|e_i)(e_j|$. Let $\theta_1,\dots\theta_n$ be real numbers and set
$\Theta:=\theta_1 E_{11}+\cdots+\theta_n E_{nn}.$
 For $A\in B(\cc^n)$ define
\beq
M(A):=\i [\Theta,A]-\frac{1}{2}\sum_jm_{jj}[E_{jj},A]_++
\sum_{i\neq j} m_{ij} E_{ij} AE_{ji}.\label{wcl6}\eeq
Then $M$ is the generator of a
Markov c.p. semigroup on $B(\cc^n)$.
The algebra $\cP$ generated by $E_{11},\dots, E_{nn}$ is preserved by
 $\e^{tM}$ and naturally
isomorphic to $\cc^n$. Under this identification, $M\Big|_\cP$ equals $m$.
 \label{dbc2}\eet

\subsection{Invariant c.p semigroups.}
\label{Invariant c.p semigroups}

Let $K$ be a self-adjoint operator on $\cK$. Let $M$ be the
  generator of a c.p. semigroup
 on $\cK$. We say that $M$ is $K$-invariant iff
\beq M(A)=\e^{-\i tK}M(\e^{\i tK}A\e^{-\i tK})\e^{\i tK},\ \ \  t\in\rr.\eeq
We will see later on that c.p. semigroups obtained in the weak coupling limit
are  always  $K$-invariant with respect the Hamiltonian of the small system.

Note that $M$ can be split in a canonical way into $M=\i[\Theta,\cdot]+M_\d$,
where $M_\d$ is its purely dissipative part.  $M$ is $K$-invariant iff
$[\Theta,K]=0$ and $
M_\d$ is $K$-invariant. Thus in what follows it is enough to restrict
ourselves to the purely dissipative case.

The following two theorems extend Theorem \ref{dbc1} and
\ref{dbc2}.

\bet Consider the set-up of Theorem \ref{dbc1}. Suppose in addition that $K$
is a self-adjoint operator on $\cK$ with the  eigenvalues
$k_1,\dots,k_n$  and
$P_j=1_{k_j}(K)$. Let
 $M$
be $K$-invariant.  Then the algebra $\cP$ is preserved by $\e^{tM}$ (and hence
the conclusion of Theorem \ref{dbc1} holds).\eet

\bet Consider the set-up of Theorem \ref{dbc2}. If
$k_1,\dots,k_n$ are real and $
K:=k_1 E_{11}+\cdots+k_n E_{nn}$, then  $M$ is $K$-invariant.\eet

The following theorem describes the $K$-invariance on the level of a
Lindblad form. We restrict ourselves to the Markov case.

\bet
Let $\nu\in B(\cK,\cK{\otimes}\frh)$ and let $Y$ be
a self-adjoint operator on $\frh$
such that
\begin{eqnarray}
M(A)&=&- \frac12 [\nu^*\nu,A]_++\nu^*A\otimes 1\nu,\label{wcl1}\\
\nu K&=&(K{\otimes}1+1{\otimes}Y)\nu.\label{wcl2}
\end{eqnarray}
Then $M$ is the generator of a $K$-invariant
purely dissipative Markov c.p. semigroup.
\label{wcl3}\eet

\Proof We check that $ \nu^*\nu$ commutes with $K$. Then it is enough to
verify  that $A\mapsto
\nu^*A\otimes 1\ \nu$ is $K$-invariant. \endproof

There exists  a partial converse of Theorem \ref{wcl3}.

\bet Let  $M$ be the generator of a $K$-invariant
 purely dissipative Markov c.p. semigroup. Let $\frh,\nu$ realize its minimal
 Lindblad form (\ref{wcl1}).
Then there exists a self-adjoint operator $Y$ on $\frh$
such that (\ref{wcl2}) is true.
\label{wcl8}\eet

\Proof
By the uniqueness part of
 Theorem \ref{stinespring} there  exists a
unique unitary operator $U_t$ on $\frh$  such that
$\e^{\i tK}{\otimes} U_t\ \nu\ \e^{-\i tK}=\nu$. We easily check the
$U_t$ is a continuous 1-parameter unitary group so that
 $U_t$ can be written as $\e^{\i tY}$ for some self-adjoint $Y$.
\endproof

Note that  Theorems \ref{wcl3} and \ref{wcl8} have a clear
physical meaning.
 The operator $\nu$ is responsible for ``quantum jumps''.
The operator $Y$ describes the energy of the reservoir (or actually of the
part of the reservoir ``directly seen'' by the interaction).
The equation (\ref{wcl2}) describes  the energetic
balance in each quantum jump.

\subsection{Detailed Balance Condition.}
\label{Detailed Balance Condition}

In the literature the name Detailed Balance Condition (DBC) is given to
several related but non-equivalent concepts. In this subsection we discuss
some of the versions of the DBC  relevant in the weak coupling limit.

Some of the definitions of the DBC (both for classical and quantum systems)
involve the time reversal \cite{Ag,Ma,MaSt}. In the weak coupling limit one
does not need to introduce the time reversal, hence we will only discuss
versions of the DBC that do not involve this
operation. (See however \cite{DM} for a discussion of time-reversal in
 semigroups obtained in the weak coupling limit.)

Let us first recall the definition of the  classical
  Detailed Balance Condition.
 Let $p=(p_1,\dots,p_n)\in\cc^n$ be a vector with
$p_1,\dots,p_n>0$. Introduce the scalar product on $\cc^n$:
\beq (u|u')_p:=\sum_j \bar u_ju_j' p_j.\label{wcl5}\eeq
Let $\e^{tm}$ be a classical Markov semigroup
 on $\cc^n$.
We  say that $m$ satisfies the Detailed Balance Condition for $p$ iff $m$ is
self-adjoint for $(\cdot|\cdot)_p$.

Let us now consider the quantum case.
Let $\rho$ be a nondegenerate density matrix. As usual, we assume that $\cK$
 is finite
dimensional. On $B(\cK)$ we introduce
the
 scalar product
\beq (A|B)_\rho:=\Tr \rho^{ 1/2 } A^*\rho^{ 1/2 }B.\label{wcl7}\eeq

Let $M$ be the generator of a c.p. semigroup on $B(\cK)$.
 Recall that it can be uniquely
represented as
\[M=\i [\Theta,\cdot]+M_\d,\]
where $M_\d$ is its purely dissipative part
and  $\i[\Theta,\cdot]$ its Hamiltonian
part.
We say that $M$ satisfies the Detailed Balance Condition (or DBC)
for $\rho$ iff
 $M_\d$ is  self-adjoint and
$\i[\Theta,\cdot]$ is  anti-self-adjoint for $(\cdot|\cdot)_\rho$.

Note that $M$
 satisfies the DBC for $\rho$ iff $[\Theta,\rho]=0$ and
$M_\d$ satisfies the DBC for $\rho$.
Therefore, in our further analysis we will often restrict ourselves to the
 purely dissipative case.

We believe that in the quantum finite
dimensional case  the above definition of the DBC is the
most natural. It was used e.g. in
\cite{DF1} under the name of the {\em standard Detailed Balance Condition}.

A similar but different definition  of the DBC can be found in
 \cite{FGKV,Al1}.
Its only difference is
the replacement of  the scalar product
$(\cdot|\cdot)_\rho$ given in (\ref{wcl7}) with
\beq
\Tr \rho A^*B.\label{wcl9}\eeq
Note that if $M$ is $K$-invariant and $\rho$ is a function of $K$, then both
 definitions are equivalent.


   The weak coupling limit applied to a small system with a Hamiltonian $K$
interacting with  a thermal reservoir at some fixed
temperature $\beta$ always yields a Markov c.p. semigroup that is
 $K$-invariant
and satisfies the DBC for $\rho=\e^{-\beta K}/\Tr
\e^{\beta K}$;
see e.g. \cite{LeSp,DF1} and Subsect
 \ref{Weak coupling limit  for thermal reservoirs}.

There exists a close relationship between the classical and quantum DBC.

\bet Consider the set-up of Theorem \ref{dbc1}. Let $\rho$ be a density matrix
on $\cK$ with the eigenvalues $p_1,\dots,p_n$  and let  $P_j$ equal the
spectral projections of $\rho$ for the eigenvalue $p_j$.
If $M$ satisfies the DBC for $\rho$, either in the sense of
 (\ref{wcl7}) or  in the sense of  (\ref{wcl9}), then the classical
Markov semigroup $\e^{tM}\Big|_\cP$  satisfies the DBC for
$p=(p_1,\dots,p_n)$.
\eet

\bet Consider the set-up of Theorem \ref{dbc2}.
 Let $\e^{tm}$ satisfy the classical DBC for
$p=(p_1,\dots,p_n)$.
Then $M$ defined by (\ref{wcl6})
satisfies both quantum versions of the  DBC for $\rho:=p_1 E_{11}+\dots+p_n
 E_{nn}$.
\eet

 The following theorem  describes the DBC for
 $K$-invariant generators on the level of their
Lindblad form. It is an extension of   Theorem \ref{wcl3}. (Note that
 (\ref{wcl1a}), (\ref{wcl2a}) are identical to
 (\ref{wcl1}), (\ref{wcl2}) of Theorem  \ref{wcl3}).

\bet
Let $\nu\in B(\cK,\cK{\otimes}\frh)$ and $Y$ a self-adjoint operator on $\frh$
such that
\begin{eqnarray}
M(A)&=&- \frac12 [\nu^*\nu,A]_++\nu^*A\otimes 1\nu,\label{wcl1a}\\
\nu K&=&(K{\otimes}1+1{\otimes}Y)\ \nu,\label{wcl2a}\\
\Tr_\frh\nu A\nu^*&=&\nu^* \ A{\otimes}\e^{-\beta Y}\ \nu.\label{condi2}
\end{eqnarray}
Then $M$ is the generator of a $K$-invariant
purely dissipative Markov c.p. semigroup satisfying the DBC for $\rho:=
\e^{-\beta
  K}/\Tr\e^{-\beta K}$.
\label{wcl3a}\eet

\Proof
 It follows from  (\ref{wcl2})
 that $\nu^*\nu$ commutes with $\e^{-\beta K/2}$. Hence $[\nu^*\nu,\cdot]_+$ is
 self-adjoint for $(\cdot|\cdot)_\rho$.

If $M$ is a map on $B(\cK)$, then $M^{*\rho}$ will denote the adjoint for this
scalar product. Let $M_1(A)=\nu^*\ A{\otimes}1\ \nu$.
We compute:
\begin{eqnarray}
M_1^{*\rho}(A)&=&
\Tr_\frh\e^{\beta K/2}{\otimes}1\ \nu\ \e^{-\beta K/2}A\e^{-\beta K/2}\ \nu^*\
\e^{\beta K/2}{\otimes} 1\nonumber\\
&=&
\Tr_\frh\e^{\beta K/2}\nu^*\ (\e^{-\beta K/2}A\e^{-\beta K/2}{\otimes}\e^{-\beta Y})\  \nu\
\e^{\beta K/2}\label{whereas1}
\\&=&\nu^*\ A{\otimes}1\nu=M_1(A).\label{whereas2}
\end{eqnarray}
In (\ref{whereas1}) and  (\ref{whereas2}) we used (\ref{condi2}) and
(\ref{wcl2a}) respectively.
\endproof

It is possible to replace the condition (\ref{condi2})
with a different condition (\ref{condi3}).
Note that whereas  (\ref{condi2}) is quadratic in $\nu$,
 (\ref{condi3}) is linear in $\nu$.

\bet Suppose that $\epsilon$ is an antiunitary operator on $\frh$ such that
\beq(\phi{\otimes}w|\nu\psi)=(\nu\phi|\psi\otimes\e^{-\beta Y/2}\epsilon w),\ \
\phi,\psi\in\cK,\ \ w\in\frh.\label{condi3}\eeq
Then (\ref{condi2}) holds.
\label{wcl3aa}\eet

\Proof It is sufficient to assume that $A=|\psi)(\psi|$ for some
 $\psi\in \cK$. Let
$\phi\in\cK$. Let $\{w_i\ |\ i\in I\}$ be an orthonormal basis in $\frh$.
Then
\begin{eqnarray*}
\Tr_\frh(\phi|\nu A\nu^*\phi)&=&
\sum(\phi{\otimes}w_i|\nu\psi)(\nu\psi|\phi{\otimes}w_i)\\[2ex]
&=&
\sum(\nu\phi|\psi{\otimes}\e^{-\beta  Y/2}\epsilon w_i)
(\psi{\otimes}\e^{-\beta Y/2}\epsilon w_i|\nu\phi)\\[2ex]
&=&\Big(\nu\phi\Big|\ |\psi)(\psi|{\otimes}\e^{-\beta Y}\ \nu\phi\Big)\\[2ex]
&=&(\phi|\nu^*\ A{\otimes}\e^{-\beta Y}\ \nu\phi).
\end{eqnarray*}
\endproof

There exists an extension of Theorem
\ref{wcl8} to the Detailed Balance Condition. It can be viewed as a
 partial converse of Theorems \ref{wcl3a} and \ref{wcl3aa}:

\bet Let  $M$ be the generator of a $K$-invariant
 purely dissipative Markov c.p. semigroup satisfying the DBC for
 $\e^{-\beta
  K}/\Tr\e^{-\beta K}$.
 Let $\frh,\nu$ realize its minimal
 Lindblad form (\ref{wcl1a}). Let a self-adjoint operator $Y$ on $\frh$
satisfy (\ref{wcl2a}).
Then (\ref{condi2}) is true and there exists
 a unique antiunitary operator $\epsilon$ on $\frh$
such that
 (\ref{condi3}) holds. Besides, $\epsilon Y\epsilon=-Y$ and $\epsilon^2=1$.
\label{wcl8b}\eet

\Proof
\noindent{\bf Step 1.} By the proof of Theorem \ref{wcl3a}, the DBC for
 $\e^{-\beta
  K}/\Tr\e^{-\beta K}$ together with
 (\ref{wcl2a}) imply (\ref{condi2}).

\noindent
{\bf Step 2.} The next step is to prove that  (\ref{condi2}) and
 (\ref{miniu}) imply the existence of an antiunitary
$\epsilon$ on $\frh$ satisfying (\ref{condi3}).

Identify $\frh$ with $\cc^n$, so that we have a complex  conjugation $w\mapsto
\overline w$ in $\frh$. We can assume that $Y$ is diagonal, so that
$\overline{Y\overline{ w}}=Y w$, $w\in\frh$.
Define  $\nu^\star$  by
\beq(\phi{\otimes}w|\nu\psi)=(\nu^\star\phi|\psi\otimes \bar w),\ \
\phi,\psi\in\cK,\ \ w\in\frh.\label{condi7}\eeq
(Note that $\star$ is a different star from $*$ denoting the Hermitian
conjugation, see \cite{DF1}).
We can rewrite (\ref{condi2}) as
\begin{eqnarray}
\nu^{\star*}\ A{\otimes}1\ \nu^{\star}&=&
\nu^{*}\ 1{\otimes}\e^{-\beta Y/2}
 \ (A{\otimes}1)\  1{\otimes}\e^{-\beta Y/2}\ \nu.\label{condi4}
\end{eqnarray}

(\ref{condi4}) defines a c.p. map. By the uniqueness part of
Theorem  \ref{stinespring} and  (\ref{miniu}),
we obtain the existence of a unitary map $U$ on $\frh$ such that
$\nu^{\star}=1{\otimes}U\e^{-\beta Y/2}\ \nu$. Now we set
$\epsilon w=U^*\bar w$.

\noindent{\bf Step 3.} We apply (\ref{condi2}) twice:
\[(\phi{\otimes} w|\nu\psi)=
(\nu\phi|\psi{\otimes}\e^{-\beta Y/2}\epsilon
w)=(\phi{\otimes}(\e^{-\beta Y/2}\epsilon)^2w|\nu\psi).\]
Using  (\ref{miniu}) we obtain $w=(\e^{-\beta Y/2}\epsilon)^2w$.

\noindent{\bf Step 4.}
Finally applying  (\ref{condi2}) together with
 (\ref{wcl2a}) twice we obtain
\[(\phi{\otimes} w|\nu\psi)=
(\nu\e^{\beta K/2}\phi|\e^{-\beta K/2}\psi{\otimes}\epsilon
w)=(\phi{\otimes}\epsilon^2w|\nu\psi).\]
Thus with help of  (\ref{miniu}) we get
 $w=\epsilon^2 w$. \endproof

Note that the above results show that
for c.p. Markov semigroups that are  $K$-invariant
and satisfy the DBC for
 $\e^{-\beta  K}/\Tr\e^{-\beta K}$  we naturally obtain  a certain
 algebraic structure on the ``restricted reservoir'' $\frh$
that resembles closely the famous
Tomita-Takesaki theory. The properties of
$\e^{-\beta Y}$ and
$\epsilon$ are paralel to those of the {\em modular operator} and the
{\em modular conjugations} -- the basic objects of the
Tomita-Takesaki formalism. (See also Subsection
\ref{Weak coupling limit for thermal reservoirs}).

\section{Bosonic reservoirs.}
\label{Bosonic reservoirs}

In this section we recall basic terminology related to second quantization,
see e.g. \cite{De0}.
We also
introduce {\em Pauli-Fierz operators} -- a class of models (known in
the literature under
various names) that are often used to describe realistic physical systems, see
e.g. \cite{DJ1,DJP}.
\subsection{Second quantization.}
\label{Second quantization}

Let  $\cH_\res$ be a Hilbert space describing 1-particle states. The
corresponding bosonic
Fock space is defined as
\[\Gamma_\s(\cH_\res):=\loplus_{n=0}^\infty\otimes_\s^n\cH_\res.\] The vacuum
vector is
 $\Omega=1\in\otimes_\s^0\cH_\res=\cc$.\\

If $z\in \cH_\res$, then
\[a(z)\Psi:=\sqrt{n}(z|{\otimes}1^{(n-1){\otimes}}\Psi\in
\otimes_\s^{n-1}\cH_\res
,\ \ \ \Psi\in \otimes_\s^n\cH_\res,\]
is called
 the annihilation operator of $z$ and $a^*(z):=a(z)^*$ is the corresponding
creation operator. They are closable operators on $
\Gamma_\s(\cH_\res)$.

For an operator $q$ on $\cH_\res $ we define the operator $\Gamma(q)$ on
$\Gamma_\s(\cH_\res )$ by
\beq\Gamma(q)\Big|_{\otimes_\s^n\cH_\res }=q\otimes\cdots\otimes
q.\label{qua2}\eeq

For an operator $h$ on $\cH_\res $ we define the  operator $\d\Gamma(h)$ on
$\Gamma_\s(\cH_\res )$ by
\[\d\Gamma(h)\Big|_{\otimes_\s^n\cH_\res }=h\otimes1^{(n-1)\otimes}+\cdots
1^{(n-1)\otimes}\otimes h.\]

Note the identity $\Gamma(\e^{\i t h})=\e^{\i t\d\Gamma(h)}$.

\subsection{Coupling  to a bosonic reservoir.}
\label{Coupling a quantum system to a bosonic resrvoir}

Let $\cK$ be a finite dimensional Hilbert space.
We imagine that it describes a small quantum system interacting with a bosonic
reservoir described by the Fock space $\Gamma_\s(\cH_\res)$.
The coupled system is described by the Hilbert space
$\cH:=\cK\otimes\Gamma_\s(\cH_\res)$.

Let $V\in B(\cK,\cK\otimes\cH_\res)$. For
$ \Psi\in \cK\otimes\otimes_\s^n\cH_\res$ we set
\[a(V)\Psi:=\sqrt{n}V^*{\otimes}1^{(n-1){\otimes}}\Psi\in
\cK\otimes\otimes_\s^{n-1}\cH_\res.
\]
$a(V)$ is called the annihilation operator of $V$ and
$a^*(V):=a(V)^*$ the corresponding creation operator. They are
closable operators on $\cK\otimes \Gamma_\s(\cH_\res)$.
Note in particular that if $V$ is written in the form
$\sum_jV_j{\otimes}|b_j)$ (which is always possible),  then
\[a^*(V)=\sum_jV_j\otimes a^*(b_j),\ \  a(V)=V_j^*\otimes a(b_j),\]
where $a^*(b_j)$, $a(b_j)$ are the usual creation/annihilation
operators introduced in the previous subsection.

%



The following class of operators plays the central role in our article:
\begin{eqnarray}\label{pauli}
H_\lambda&=&
K\otimes1+1\otimes \d\Gamma(H_\res)+\lambda( a^*(V)+ a(V)).\end{eqnarray}
Here $K$ is a self-adjoint operator describing the free dynamics  of the small
system,
$\d\Gamma(H_\res)$
 describes the free dynamics of
the reservoir
and $ a^*(V)$/$ a(V)$,
for some $V\in B(\cK,\cK\otimes\cH_\res)$, describe the interaction.
Operators of the form \ref{pauli}
will be called {\em  Pauli-Fierz operators}.

Note that operators of the form (\ref{pauli}) or similar are very
common in the physics literature and are believed to give an
approximate description of realistic physical systems in many
circumstances (e.g. an atom interacting with radiation in
the dipole approximation), see e.g. \cite{DJ1}.

\subsection{Thermal reservoirs.}
\label{Weak coupling limit for thermal reservoirs}

In this subsection we will discuss thermal reservoirs.
 We fix a positive number $\beta$ having the interpretation of the
 inverse temperature.

 Recall that the free Hamiltonian is
 $H_0:=K{\otimes}1+1{\otimes}\d\Gamma(H_\res)$. To have a simpler formula for
 the Gibbs state of the small system we assume that
 $\Tr\e^{-\beta K}=1$. We set
\begin{eqnarray*}
\tau_{t}(C)&:=&\e^{\i
  tH_0}C\e^{-\i tH_0},\\
\omega_{\beta}(C)&:=&\Tr\e^{-\beta K} {\otimes}|\Omega)(\Omega|\
C
,\ \ C\in B(\cH).\end{eqnarray*}

\bet
The following are equivalent:

\ben\item  For any $D_1,D_2,D_1', D_2'\in B(\cK)$ and
 \begin{eqnarray*}
B_j&:=&D_j{\otimes}1 \left(a^*(V)+a(V)\right)\ D'_j \otimes  1,\ \
j=1,2,
\end{eqnarray*}
 and for any $t\in \rr$
we have
\beq
\omega_{\beta}(\tau_{t}(B_1)B_2)=\omega_{\beta}\left(B_2\tau_{t+\i\beta}(B_1)\right).\label{kms}\eeq
\item
For any function $f$ on the spectrum of $\sp H_\res$ and
$A \in \mathcal{B}(\mathcal{K})$, we have \beq \Tr_{\cH_\res}
1{\otimes}\bar f(-H_\res)\ V \ A\ V^* = V^*\ A{\otimes}
 \e^{-\beta H_\res}
f(H_\res)\ V
.\label{kms1}\eeq
\een\label{kms4}\eet

\Proof
The left hand side of (\ref{kms}) equals
\[\Tr \e^{-\beta K+\i tK}D_1 V^*(D_1^{\prime}\e^{-\i tK}
D_2{\otimes} \e^{-\i tH_\res})
V D_2^\prime.\]
 The right hand side of (\ref{kms})
equals
\[\Tr D_2 V^*(D_2^\prime\e^{-\beta K+\i tK} D_1{\otimes}\e^{(-\beta +\i
 t)H_\res} )
   V
 D_1^{\prime}\e^{-\i tK}.\]
Now we set $A_1:=D_2^\prime \e^{-\beta K+\i tK}D_1$,
 $A_2:=D_1^{\prime}\e^{-\i tK}D_2$, and use
 the cyclicity of the trace. We
 obtain
\[\Tr A_2{\otimes}\e^{-\i tH_\res}\ V A_1\ V^* =
\Tr  A_2 V^*
A_1{\otimes}\e^{-\beta H_\res+\i tH_\res}\ V.\]
By the Fourier transformation we get
\[\Tr A_2{\otimes} \bar f(-H_\res)\ V A_1\ V^*=
\Tr A_2 V^* A_1{\otimes}\e^{-\beta H_\res}  f(H_\res) \ V
.\]
This implies (\ref{kms1}).
\endproof

We will say  that the reservoir is thermal at the inverse
temperature $\beta$ iff the conditions of Theorem \ref{kms4} are
true.

 (\ref{kms}) is just the $\beta$-KMS
condition
 for the state $\omega_\beta$, the dynamics $\tau$ and appropriate operators.
Note that (\ref{kms}) is satisfied for Pauli-Fierz semi-Liouvilleans
constructed with help of the Araki-Woods representations of the CCR, where we
use the terminology of \cite{DJP,De0}.
Theorem \ref{kms4} describes
 a substitute of the  KMS condition without
 invoking explicitly  operator algebras.

The KMS condition is
 closely related to the Tomita-Takesaki theory. One of the objects introduced
 in this theory is the modular conjugation. It turns out that the
 set-up of Theorem \ref{kms4} is sufficient to introduce
 a substitute for the modular conjugation without talking about operator
 algebras.

Define
\[\cH_{\tilde{\res}}:=\{(\phi|{\otimes}f(H_\res)\  V\psi\ :\ \phi,\psi\in\cK,\ f\in
C_\c(\rr)\}^\cl.\]
($\cl$ denotes the closure).  Clearly, $\cH_{\tilde{\res}}$ is a subspace of
$\cH_\res$ invariant with respect to the 1-particle reservoir Liouvillean
$H_\res$. It describes the part of $\cH_\res$ that is coupled to the small
system. Let $H_{\tilde{\res}}$ denote the operator $H_\res$ restricted to the space
$\cH_{\tilde{\res}}$.

\bet Suppose that the reservoir is thermal at inverse temperature $\beta$. Then
there exist a unique antiunitary operator $\epsilon_{\tilde{\res}}$ on
 $\cH_{\tilde{\res}}$ such that
\beq
(\phi{\otimes} w| V\psi)=
(V\phi|\psi{\otimes}\e^{-\beta H_{\tilde{\res}}}\epsilon_{\tilde{\res}} w).
\eeq It satisfies $\epsilon_{\tilde{\res}}^2=1$ and
 $\epsilon_{\tilde{\res}} H_{\tilde{\res}} \epsilon_{\tilde{\res}}=-H_{\tilde{\res}}$.
\eet

\Proof For $f\in C_\c(\rr)$, $\phi,\psi\in\cK$, we set
\[\epsilon_{\tilde{\res}}\left(
(\phi|{\otimes}\e^{-\beta H_{\tilde{\res}}/2}f(H_{\tilde{\res}})\ V\psi\right)
:=
(\psi|{\otimes}
\bar f(-H_{\tilde{\res}}) V\phi.\]
 (\ref{kms1})
implies that $\epsilon_{\tilde{\res}}$ is a well defined antiunitary map. \endproof

\section{Quantum Langevin dynamics.}
\label{Quantum Langevin dynamics}

 Suppose that we are given
a
c.p. Markov semigroup $\e^{tM}$ on $B(\cK)$.
We will describe a certain class of  self-adjoint operators $Z$ on a larger
Hilbert space such that
$\e^{-\i tZ}\cdot\e^{\i tZ}$ is a dilation on $\e^{tM}$.
We will use
 the name {\em quantum Langevin }(or {\em stochastic}) {\em dynamics}
for $\e^{-\i tZ}\cdot\e^{\i tZ}$.
The unitary group $\e^{-\i tZ}$ will be called
a {\em Langevin }(or {\em stochastic}) {\em Schr\"odinger dynamics}.

In Subsection \ref{Linear noises}
we will restrict ourselves to a subclass of quantum Langevin dynamics
involving only the so-called linear noises. Actually, at present our
 results on the extended
weak coupling limit are limited only to them.

In Subsection \ref{Quadratic noises} we will describe a more general class
of quantum Langevin dynamics, which also involve {\em quadratic noises}.
Our construction involving quadratic noises is related to the
 operator-theoretic
 approach of Chebotarev \cite{Ch,ChR}, and especially of Gregoratti
\cite{Gr}.

 We
expect that our approach to the extended weak coupling limit can be improved
to cover also
 this larger class. Within the approach of \cite{AFL} there exist partial
results in this direction \cite{Go}.

The history of the discovery of quantum Langevin dynamics is quite involved.
The construction can be traced back to \cite{AFLe}, and especially \cite{HP}
where the quantum stochastic calculus was introduced. But apparently  only
in  \cite{Fr} and \cite{Maa} it was independently realized that
this leads to a dilation of  Markov c.p. semigroups.
 Let us also mention \cite{At,Me,Fa} for more recent
presentations of the quantum stochastic calculus.

\subsection{Linear noises.}\label{Linear noises}

Apart from a c.p. Markov semigroup
$\e^{tM}$ let us fix some additional data.
 More precisely, we fix
an operator $\Upsilon$, an auxiliary Hilbert space $\frh$ and an
 operator $\nu$ from $\cK$ to $\cK\otimes\frh$ such that
\[-\i\Upsilon+\i\Upsilon^*=-\nu^*\nu\]
and $M$ is given by \begin{equation*}\label{lindi4} M(A)= -\i
(\Upsilon A - A \Upsilon^*) +  \nu^*\ A{\otimes}1\ \nu,   \qquad A
\in B(\cK).\end{equation*} In other words, we fix a concrete
Lindblad form of $M$.

Introduce the Hilbert space
 $\cZ_\res:=\frh\otimes L^2(\rr)$. The enlarged  Hilbert space is
$\cZ:=\cK\otimes \Gamma_\s(\cZ_\res)$.

Let
$Z_\res$ be the operator of
multiplication by the variable $x$ on $L^2(\rr)$.
Let  $(1|$, $|1)$  be defined as in (\ref{111}).

 We choose a basis $(b_j)$ in $\frh$, so that we can write
\beq \label{basis chosen}\nu=\sum\nu_j\otimes|b_j).\eeq (Note that
at the end the construction will not depend on the choice of a
basis). Set
\begin{eqnarray*} \nu_{j}^+=\nu_j,&&
\nu_{j}^-=\nu_j^*.
\end{eqnarray*}

For $t\geq0$ we define the quadratic form
\begin{eqnarray*}
U_t&:=&
\e^{-\i t\d\Gamma(Z_\res)} \sum_{n=0}^\infty
\mathop{\int}\limits_{t\geq t_n\geq\cdots\geq t_1\geq0}
\d t_n\cdots\d t_1\\
&&\times (2\pi)^{-\frac{n}{2}}\sum_{j_1,\dots,j_n}
\ \sum_{\epsilon_1,\dots,\epsilon_n\in\{+,-\}}\nonumber \\
&&\times (-\i)^n\e^{-\i(t-t_n)\Upsilon}
\nu_{j_n}^{\epsilon_n}\e^{-\i(t_n-t_{n-1})\Upsilon}\cdots
\nu_{j_1}^{\epsilon_1}\e^{-\i(t_1-0)\Upsilon}\nonumber\\
&&\times\prod_{k=1,\dots,n:\  \ \epsilon_k=+} a^*( \e^{\i t_k Z_R}
b_{j_k} \otimes |1))\\&&\times \prod_{k'=1,\dots,n:\  \
\epsilon_{k'}=-}
a(  \e^{\i t_{k'} Z_R} b_{j_{k'}}  \otimes  |1));\nonumber\\
U_{-t}&:=&U_t^*.
\end{eqnarray*}

We will denote by $I_\cK$ the embedding of $\cK\simeq\cK\otimes\Omega$ in $\cZ$.
\bet
$U_t$ extends to
a strongly continuous unitary group on $\cZ$ such that
\begin{eqnarray*}
I_\cK^*U_tI_\cK&=&\e^{-\i t\Upsilon},\\
I_\cK^*U_t\  A\otimes1\ U_{-t}I_\cK&=&\e^{tM}(A).
\end{eqnarray*}
Thus $U_t$ is a unitary dilation of $\e^{-\i t\Upsilon}$, and
$U_t\cdot U_t^*$ is a dilation of $\e^{tM}$. \eet

As every strongly continuous unitary group, $U_t$ can be written as $\e^{-\i
  tZ}$ for a certain self-adjoint operator $Z$.
Note that
formally (and also rigorously with an appropriate regularization)
\begin{eqnarray*}
Z&=&\frac12(\Upsilon+\Upsilon^*)+\d
\Gamma(Z_\res)\\
&&+(2\pi)^{-\frac{1}{2}}
a^*\left(\nu\otimes|1)\right)+(2\pi)^{-\frac{1}{2}}a\left(\nu\otimes|1)\right).
\end{eqnarray*}
Thus $Z$ has the form of a Pauli-Fierz operator with a rather singular interaction.

Let us present an alternative variation of the above construction, which
 is actually closer to what can be found in the literature.
Let $\cF$ be the Fourier transformation on $\cZ_\res=
\frh\otimes L^2(\rr)$ defined as in (\ref{fou}).
  The operator $Z$  transformed by  $1_\cK{\otimes}\Gamma(\cF)$  will be
 denoted by
\begin{eqnarray}
\hat Z&:= &1_\cK{\otimes}\Gamma(\cF)\ Z\ 1_\cK{\otimes}\Gamma(\cF^*)
\label{expoa1}.
\end{eqnarray}
It equals
\begin{eqnarray}\nonumber\hat Z&=&
\frac12(\Upsilon+\Upsilon^*){\otimes}1 + 1{\otimes}\d\Gamma(D_\tau)\\[2ex]
&& +
   a\left( \nu\otimes|\delta_0) \right)
+a^*\left( \nu\otimes|\delta_0)\right),\nonumber
\end{eqnarray}
where $\delta_0$, $D_\tau$ are defined as in (\ref{defo1}), (\ref{defo2}).

 Similarly to the operator of
Section \ref{Dilations of contractive semigroups} denoted with the same
symbol, the operator $\hat Z$ (as well as $Z$)
has a number of intriguing properties.
Let us describe one of them.

Let $\cD_0:=\frh\otimes H^1(\rr)$. (Recall that $H^1(\rr)$ is the
first Sobolev space). Let $\Gammal_\s(\cD_0)$,
 denote the corresponding algebraic Fock
space and  $\cD_1:=\cK\otimes\Gammal_\s(\cD_0)$.
 Introduce the (non-self-adjoint) sesquilinear form
\begin{eqnarray}
\hat Z^+& =&
\Upsilon{\otimes}1  + 1{\otimes}\d\Gamma(D_\tau)\\[2ex]&&+
a\left( \nu\otimes|\delta_0 )\right)
+a^*\left( \nu\otimes|\delta_0)\right)
 .\nonumber
\end{eqnarray}

Let $\psi,\psi' \in \cD_1$. Then \beq \lim_{t \downarrow 0}
\frac{1}{ t}( \psi| (\e^{-\i t\hat Z}-1)  \psi') = -\i( \psi|
\hat{Z}^+  \psi'). \label{kah1}\eeq  Thus it seems that $\hat
Z^+=\hat Z$, which is true only if $\Upsilon$ is self-adjoint and
hence there are no off-diagonal terms in $Z$. Clearly, the
explanation of the above paradox is similar as in Subsect.
\ref{Dilations of contractive semigroups}: $(\psi | \e^{-\i t\hat
Z}\psi')$ is not differentiable at zero. This is related to the
fact that
  $\psi,\psi'$ do not belong to $\Dom Z$.
 Thus $\hat Z^+$ can again be called
a {\em false form}.

In the literature, the Langevin Schr\"odinger dynamics
$ \e^{-\i t\hat Z }$ is usually introduced  through the so-called
 {\em Langevin }(or {\em  stochastic})
 {\em Schr\"odinger equation}
satisfied by
 \beq \hat W(t):=
\e^{\i t\d \Gamma (D_\tau) } \e^{-\i t\hat Z }. \eeq To write this
equation recall the decomposition \eqref{basis chosen} and note
that. Then, in the sense of quadratic forms on $\cD_1$, we have
\beq \i \frac{\d}{\d t}\hat W(t) = \big(\Upsilon {\otimes}1+  a^*
\left(\nu \otimes|\delta_t ) \right) \big)\hat W(t) +\sum_j
\nu^*_j\hat W(t) a \left(b_j \otimes|\delta_t ) \right).
\label{schr}\eeq

Note that $a\left( \nu\otimes|\delta_0 )\right)$ and
$a^*\left( \nu\otimes|\delta_0)\right)$ appearing in $\hat Z$ and $\hat Z^+$
 are quantum analogs of  a classical
{\em white noise}. They are ``localized'' at
$\tau=0$. Besides,
they are (formally) given by a linear expression in terms of
creation/annihilation operators.  Therefore, they are often called {\em
  linear quantum
  noises}.

\subsection{Quadratic noises.}
\label{Quadratic noises}

This subsection is outside of the main line of this article. It is closely
related to Subsect. \ref{``Toy quadratic noises''}.
It is not needed for the description of the weak coupling limit, as given in
the next section.

 Clearly,
$\Psi\in\cK\otimes\left(\otimes_\s^n\frh\otimes L^2(\rr)\right)
\simeq \cK\otimes\left(\otimes_\s^n L^2(\rr,\frh)\right)$
 can be identified with a function
$\Psi(\tau_1,\dots,\tau_n)$ with values in $\cK\otimes(\otimes^n\frh)$ and the
arguments satisfying $\tau_1<\dots<\tau_n$.

Let $S$ be a unitary operator on $\cK\otimes\frh$. Let $S_{(j)}$ be this
operator acting on
$\cK\otimes\otimes^n\frh$, where it is applied to the $j$'th ``leg'' of the
tensor product $\otimes^n\frh$. We define
an operator $\Lambda(S)$ on $\cK\otimes\left(\otimes_\s^n
L^2(\rr,\frh)\right)$ as
follows: If
$\tau_1<\cdots<\tau_k<0<\tau_{k+1}<\cdots<
\tau_n$, then
\[\left(\Lambda(S)\Psi\right)(\tau_1,\dots,\tau_n)
:=S_{(k+1)}\cdots S_{(n)}\Psi(\tau_1,\dots,\tau_n).\]
Clearly, $\Lambda(S)$ is a unitary operator. If $\cK=\cc$, then it
coincides with  $\Gamma(\gamma(S))$, where $\gamma(S)$ was defined in
(\ref{qua1}) and $\Gamma$ is the {\em functor of the second quantization}
defined in (\ref{qua2}).

Introduce the operator $\hat Z_{S,0}$
 on $\cK{\otimes}\Gamma_\s(\cZ_\res)$ by
\beq
\hat Z_{S,0}:=
\Upsilon+\Lambda(S)^*1{\otimes}\d\Gamma(D_\tau)\Lambda(S).\label{singu}\eeq
The operator (\ref{singu}) is very singular and  contains a ``delta
 interaction at $\tau=0$''.

Let us now define the dynamics $\hat U_{S,t}$ that generalizes
$\hat U_t$. Let $S_{ij}\in B(\cK)$ be defined by \beq
S=\sum_{i,j}S_{ij}{\otimes}|b_i)(b_j|.\eeq Set
\begin{eqnarray*}
\nu_{S,j}^{+}=\nu_j,& &
\nu_{S,j}^-=\sum_i\nu_i^*S_{ij}.
\end{eqnarray*}

Then we introduce the quadratic form
\begin{eqnarray*}
\hat U_{S,t}&:=&
\sum_{n=0}^\infty
\mathop{\int}\limits_{t\geq t_n\geq\cdots\geq t_1\geq0}
\d t_n\cdots\d t_1\\
&&\sum_{j_1,\dots,j_n}
\ \sum_{\epsilon_1,\dots,\epsilon_n\in\{+,-\}}\nonumber \\
&& \times (-\i)^n\prod_{k=1,\dots,n:\  \ \epsilon_k=+} a^*(
   b_{j_k}  \otimes |\delta_{t_k-t})) \\&& \e^{-\i(t-t_n)\hat Z_{S,0}}
\nu_{S,j_n}^{\epsilon_n}{\otimes}1\e^{-\i(t_n-t_{n-1})\hat
Z_{S,0}} \cdots
\nu_{S,j_1}^{\epsilon_1}{\otimes}1\e^{-\i(t_1-0)\hat
  Z_{S,0}}\nonumber
\\&&\times \prod_{k'=1,\dots,n:\  \
\epsilon_{k'}=-}
a(  b_{j_{k'}} \otimes |\delta_{t_{k'}}));\nonumber\\
\hat U_{S,-t}&:=&\hat U_{S,t}^*.
\end{eqnarray*}
One can check that $\hat U_{S,t}$ extends to a strongly continuous unitary
group. Therefore, one can define a self-adjoint operator $\hat Z_S$ such that $
\hat U_{S,t}=\e^{-\i t\hat Z_S}$. It satisfies
\begin{eqnarray*}
I_\cK^*\hat U_{S,t}I_\cK&=&\e^{-\i t\Upsilon},\\
I_\cK^*\hat U_{S,t}\  A\otimes1\ \hat U_{S,-t}I_\cK&=&\e^{tM}(A).
\end{eqnarray*}

It is awkward to write a formula for $\hat Z_S$ in terms of
creation/annihilation operators, even formally. There exists however and
alternative formalism that is  commonly used in the literature
to define
the group $\e^{-\i t\hat Z_S}$.
Let $\psi,\psi' \in \cD_1$.
Introduce the cocycle \beq \hat W_S(t):= \e^{\i t \d\Gamma( D_\tau)}
\e^{-\i t\hat Z_S}. \eeq Then, in the sense of a quadratic form on $\cD_1$, the
cocycle  satisfies the differential equation
 \begin{eqnarray} \i\frac{\d}{\d t}
\hat W_S(t)&=& \big(\Upsilon{\otimes}1+ a^*( \nu {\otimes}
|\delta_t))\big)\ \hat W_{S}(t)\\&&
+\sum_{ij}\i(1-S_{ij}){\otimes}a^*( b_i \otimes |\delta_t))\ \hat
W_{S}(t)
\ a( b_j {\otimes} |\delta_t))\\
&& +\sum_{j} \nu_{S,j}^- \ \hat W_S(t)a(b_j {\otimes} |\delta_t)).
 \label{quadratic qsde}
\end{eqnarray}
This formula is the {\em quantum Langevin (stochastic)
equation } for the cocycle $\hat W_S(t)$ in the sense of
\cite{HP,Fa,Pa,At,Maa,Fr,Bar,Me}, which includes all three kinds
of noises. In the literature, the dilation  $ \e^{-\i t\hat Z_S}$
is usually introduced through a version of (\ref{quadratic qsde}).

\subsection{Total energy operator.}

Let us analyze  the impact of  the invariance of a c.p. semigroup on
its quantum Langevin dynamics.

Suppose now that $K$ is a self-adjoint operator on  $\cK$ and $Y$
a self-adjoint operator on $\frh$. Assume that they satisfy \beq \
\nu \ K=(K{\otimes}1+1{\otimes} Y)\nu, \ \
\left[\frac12(\Upsilon+\Upsilon^*),K\right]=0. \label{condi6b}\eeq
This implies in particular that $M$ is
 $K$-invariant.
Define the self-adjoint operator on $\cZ$
\beq E:=K{\otimes}1+1{\otimes}\d\Gamma(Y{\otimes}1).\label{tota}\eeq
Then it is easy to see that the
quantum Langevin dynamics commutes with this operator:
\beq [E,\e^{-\i tZ}]=0.\eeq
  $E$ will be called the {\em total energy operator}, which is a name
suggested by the physical interpretation that we attach to $E$.

Next we discuss the implications of the DBC of a c.p. semigroup on its quantum
 Langevin dynamics.
 We set
\begin{eqnarray*}
\sigma_{t}(C)&:=&\e^{\i
  tE}C\e^{-\i tE},\\
\omega_{\beta}(C)&:=&\Tr\e^{-\beta K} {\otimes}|\Omega)(\Omega|\
C/\Tr\e^{-\beta K}
,\ \ C\in B(\cZ).\end{eqnarray*}
We will see that the DBC for $\e^{-\beta K}/\Tr\e^{-\beta K}$ is related
to  a version of the
  $\beta$-KMS condition for the dynamics $\sigma_t$
 and the state $\omega_\beta$.

\bet Assume (\ref{tota}).
Then the following statements are equivalent:

\ben\item  For any $D_1,D_2,D_1', D_2'\in B(\cK)$, $f_1,f_2\in L^2(\rr)$
and
 \begin{eqnarray*}
B_j&:=&D_j{\otimes}1\
\big(a^*(\nu{\otimes}|f_j))+a(\nu{\otimes}|f_j)) \big)\ D'_j{
\otimes} 1 ,\ \ j=1,2.
\end{eqnarray*}
 and for any $t\in \rr$
we have
\beq
\omega_{\beta}(\sigma_{t}(B_1)B_2)
=\omega_{\beta}\left(B_2\sigma_{t+\i\beta}(B_1)\right).\label{kmsd}\eeq
\item
\begin{eqnarray}
\Tr_\frh\nu A\nu^*&=&\nu^* \ A{\otimes}\e^{-\beta Y}\ \nu
,\label{condi5d}\end{eqnarray}
(This implies in particular that  $M$ satisfies
the DBC for  $\e^{-\beta K}/\Tr\e^{-\beta K}$).
\een \eet

\section{Weak coupling limit for Pauli-Fierz operators.}
\label{Weak coupling limit for Pauli-Fierz operators}

In this section we describe the main results of this article. They are devoted
to a rather large class of Pauli-Fierz operators in
the weak coupling limit. In the first subsection we recall the well known
results about the reduced dynamics, which go back to Davies
\cite{Da1,Da2,Da3}.
 In
the second subsection we describe our results that
include the reservoir \cite{DD2}. They are  inspired by \cite{AFL}.
Finally, we discuss the case of thermal reservoirs.

\subsection{Reduced weak coupling limit.}
\label{Reduced weak coupling limit for Pauli-Fierz operators}

We consider a Pauli-Fierz operator
\begin{eqnarray*}
H_\lambda&=&K\otimes1+1\otimes \d\Gamma(H_\res)
+\lambda( a^*(V)+ a(V)).\end{eqnarray*}
We assume that $\cK$ is finite dimensional and
 for any $A\in B(\cK)$ we have
$\int \|V^* A\otimes1 \ \e^{-\i tH_0}V\|\d t<\infty.$
The following theorem is essentially a special case of a result of Davies
\cite{Da1,Da2,Da3}, see also \cite{DD2}.
\bet[Reduced weak coupling limit for Pauli-Fierz operators]
There exists a $K$-invariant Markov c.p. semigroup $\e^{tM}$ on $B(\cK)$
such that
\[\lim_{\lambda\searrow0}
\e^{-\i tK/\lambda^2} I_\cK^*\e^{\i tH_\lambda/\lambda^2}\
A\otimes1\
\e^{-\i tH_\lambda/\lambda^2}I_\cK
\e^{\i tK/\lambda^2}=\e^{tM}(A),
\]and a contractive semigroup $\e^{-\i t\Upsilon}$ on $
\cK$
 such that $[\Upsilon,K]=0$ and
\[\lim_{\lambda\searrow0}
\e^{\i tK/\lambda^2} I_\cK^*\e^{-\i tH_\lambda/\lambda^2}I_\cK
=\e^{-\i t\Upsilon}.\] If the reservoir is at inverse temperature
$\beta$, then $M$ satisfies the DBC for the state $\e^{-\beta
K}/\Tr\e^{-\beta K}$. \eet

The operator $\Upsilon\in B(\cK)$ arising in the weak coupling
limit equals
\begin{eqnarray*}
 \Upsilon &:= &
-\i\sum_{\omega}\ \sum_{k-k'=\omega}\int_0^\infty
 1_{k}(K) V^*  1_{k'}(K)\e^{-\i t (H_\res
-\omega)} V  \, 1_{k}(K) \d t.
 \end{eqnarray*}

In order to write an explicit formula for $M$ it is convenient to introduce an
additional assumption, which  anyway will be useful later on in the extended
weak coupling limit.

\begin{assumption}
Suppose that for any $\omega\in\sp K-\sp K$ there exist an open
$I_\omega\subset \rr$  and a Hilbert space $\frh_\omega$
 such that $\omega\in I_\omega$ and
\begin{eqnarray*}
\Ran 1_{I_\omega}(H_\res)&\simeq &\frh_\omega\otimes L^2(I_\omega,\d x),
\end{eqnarray*}
$1_{I_\omega}(H_\res)H_\res $ is the multiplication operator by the variable
 $x\in I_\omega$ and, for $\psi\in\cK$,
\begin{eqnarray*}
 1_{I_\omega}(H_\res) V\psi&\simeq & \int_{I_\omega}^\oplus v(x)\psi\d x.
\end{eqnarray*}
Assume that $I_\omega$ are disjoint for distinct $\omega$ and
$x\mapsto v(x)\in B(\cK,\cK{\otimes}\frh_\omega)$ is continuous at $\omega$.
\end{assumption}

Thus we assume that the reservoir 1-body Hamiltonian $H_\res$ and the
interaction $V$ are well behaved around the {\em Bohr frequencies} --
differences of eigenvalues of $K$.

Let $\frh:=\loplus_{\omega}\frh_\omega$.
We define
$\nu_\omega\in B(\cK,\cK{\otimes}\frh_\omega)$
by
\begin{eqnarray*}
\nu_\omega:=(2\pi)^{\frac{1}{2}}\sum_{\omega=k-k'}1_{k}(K)v(\omega)1_{k'}(K)\end{eqnarray*}
and $\nu\in B(\cK,\cK{\otimes}\frh)$ by
\begin{eqnarray*}
\nu:=\sum_{\omega}\nu_\omega.
\end{eqnarray*}

 Note that
\begin{eqnarray*}
\i\Upsilon-\i\Upsilon^*&=& \sum_{\omega}\
\sum_{k-k'=\omega}\int_{-\infty}^\infty
 1_{k}(K) V^*  1_{k'}(K)\e^{-\i t (H_\res
-\omega)} V  \, 1_{k}(K) \d t\\&=&
\sum_{\omega}\ \sum_{k-k'=\omega}
 1_{k}(K) v^*(\omega)  1_{k'}(K)v(\omega)  \, 1_{k}(K)\\
&=& \nu^*\nu.\end{eqnarray*}
The  generator of a c.p. Markov semigroup that arises in the reduced
weak coupling limit, called sometimes the  Davies generator, is
\begin{eqnarray}\label{davo}
 M(A)&=&
 -\i (\Upsilon A - A \Upsilon^*) +\nu^*\ A{\otimes}1\ \nu\\[2ex]
&=&-\i \left[\frac{\Upsilon +\Upsilon^*}{2},A\right]
-\frac12[A,\nu^*\nu]_+ +  \nu^* A{\otimes}1\ \nu ,   \qquad A \in
B(\cK).\nonumber\end{eqnarray}

\subsection{Energy of the reservoir in  the weak coupling limit.}

Introduce the operator $Y$ on $\frh$ by setting
\beq
 Y=\omega\ \ \ \ \hbox{on}\ \ \ \ \frh_\omega.\label{kjh}\eeq
The operator $Y$ has the interpretation of the asymptotic energy of the
restricted reservoir.

\bet \ben \item The operator $\nu$
 constructed in the weak coupling limit satisfies
\beq
\ \nu \ K=(K{\otimes}1+1{\otimes} Y)\nu.\label{condi6}\eeq
This implies in particular that $M$ is
 $K$-invariant.
\item
If the reservoir is at inverse temperature $\beta$, then $\nu$ satisfies
\begin{eqnarray}
\Tr_\frh\nu A\nu^*&=&\nu^* \ A{\otimes}\e^{-\beta Y}\ \nu
,\label{condi5}\end{eqnarray}
This implies in particular that  $M$ satisfies
the DBC for  $\e^{-\beta K}/\Tr\e^{-\beta K}$.
\een \eet

\subsection{Extended weak coupling limit.}
\label{Extended weak coupling limit for Pauli-Fierz operators}

Recall that given $(\Upsilon,\nu,\frh)$ we can define the space
$\cZ_\res$ and the Langevin Schr\"odinger dynamics $\e^{-\i tZ}$ on the
space $\cZ:=\cK\otimes\Gamma_\s(\cZ_\res)$, as in Subsect.
\ref{Linear noises}.


For $\lambda > 0$, we define the family of partial isometries
$J_{\lambda,\omega}: \frh_\omega\otimes L^2(\rr)
 \rightarrow  \frh_\omega \otimes L^2(I_\omega)\subset \cH_\res$:
\begin{equation*}
  (J_{\lambda,\omega}g_\omega)(y)=   \left\{ \begin{array}{ll}
        \frac{1}{\lambda}
        g_\omega(\frac{y-\omega}{\lambda^2}),  & \textrm{ if }   y\in
        I_\omega;
 \\
        0,  & \textrm{ if } y\in\rr\backslash I_\omega. \\ \end{array}
  \right.
\end{equation*}
We set $J_\lambda: \cZ_\res\to\cH_\res$, defined for $g=(g_\omega)$ by
\[J_\lambda g:=\sum_\omega J_{\lambda,\omega}g_\omega.\]
Note that $J_\lambda$ are partial isometries and
$\slim_{\lambda\searrow0}J_\lambda^*J_\lambda=1.$

Set $Z_0:=\d\Gamma(Z_\res)$.
The following theorem \cite{DD2} was inspired by \cite{AFL}:

\bet[Extended weak coupling limit for Pauli-Fierz operators]
\begin{eqnarray*}
&&  \s^*-\lim_{\lambda \searrow 0}\Gamma  (J^*_{\lambda} )
\e^{\i \lambda^{-2}t H_0}  \e^{-\i
\lambda^{-2}(t-t_0) H_\lambda}\e^{\i \lambda^{-2}t_0 H_0} \Gamma  (J_\lambda
)\\[4mm]
&=& \e^{\i t Z_0}\e^{-\i( t-t_0)Z }\e^{-\i t_0 Z_0}.
\end{eqnarray*}
\eet

The extended weak coupling limit can be used to describe interesting physical
properties of non-equilibrium quantum systems, see e.g. \cite{DM}.
The following corollary, which generalizes the results of \cite{Du},
 describes
the asymptotics of correlation functions for observables of the form
$ \Gamma(J_\lambda) A\Gamma(J_\lambda^*)$, where $A$ are observables on the
asymptotic space.

\begin{corollary}[Asymptotics of correlation functions]
Suppose that\\ $A_\ell,\dots,A_1\in B(\cZ)$ and
$t,t_\ell,\dots,t_1,t_0\in\rr$. Then
\vspace{1cm}
\begin{eqnarray*}&&
 \s^*-\lim_{\lambda \searrow 0}
I_\cK^*\e^{\i \lambda^{-2}t H_0}  \e^{-\i
\lambda^{-2}(t-t_\ell) H_\lambda}\e^{-\i \lambda^{-2}t_\ell H_0}
 \Gamma(J_\lambda) A_\ell\Gamma(J_\lambda^*)\\[3mm]
&&\cdots\Gamma(J_\lambda) A_1\Gamma(J_\lambda^*)
\e^{\i \lambda^{-2}t_1 H_0}  \e^{-\i
\lambda^{-2}(t_1-t_0) H_\lambda}\e^{-\i \lambda^{-2}t_0 H_0}  I_\cK\\[5mm]
&=&I_\cK^*\e^{\i t Z_0}  \e^{-\i
(t-t_\ell) Z}\e^{-\i t_\ell Z_0} A_\ell\\[3mm]
&&\cdots A_1
\e^{\i t_1 Z_0}  \e^{-\i
(t_1-t_0) Z}\e^{-\i t_0 Z_0}I_\cK.
\end{eqnarray*}
\end{corollary}

The following corollary is interesting since it describes how
reservoir Hamiltonians converge to  operators whose dynamics under
the quantum Langevin dynamics $U_{-t}\cdot U_t$ is well-studied, see e.g.\
\cite{Bar}.

\begin{corollary}[Asymptotic reservoir energies]
Consider the operator $Y:\frh \mapsto \frh$ defined in (\ref{kjh}).
The operator  $E:=K{\otimes}1 + 1{\otimes}\d \Ga (Y{\otimes}1)$
 plays the role of
``asymptotic total energy operator'', i.e.\  \beq[E, \e^{\i t
Z}]=0. \eeq Besides, for $\kappa_1,\dots,\kappa_\ell\in\rr$,
 \vspace{1cm}
\begin{eqnarray*}&&
 \s^*-\lim_{\lambda \searrow 0}
I_\cK^*\e^{\i \lambda^{-2}t H_0}  \e^{-\i \lambda^{-2}(t-t_\ell)
H_\lambda}\e^{-\i \lambda^{-2}t_\ell H_0}
 \, \e^{\i\kappa_\ell \d \Ga (H_\res) }\, \\[3mm]
&&\cdots \e^{\i\kappa_1 \d \Ga (H_\res) }\, \e^{\i \lambda^{-2}t_1 H_0}
\e^{-\i
\lambda^{-2}(t_1-t_0) H_\lambda}\e^{-\i \lambda^{-2}t_0 H_0}  I_\cK\\[5mm]
&=&I_\cK^*\e^{\i t Z_0}  \e^{-\i
(t-t_\ell) Z}\e^{-\i t_\ell Z_0} \,\e^{\i\kappa_\ell\d \Ga (Y{\otimes}1)}\, \\[3mm]
&&\cdots \, \e^{\i\kappa_1\d\Gamma(Y{\otimes}1)} \, \e^{\i t_1 Z_0}  \e^{-\i
(t_1-t_0) Z}\e^{-\i t_0 Z_0}I_\cK.
\end{eqnarray*}
\end{corollary}

\newpage
\bibliographystyle{plain}

\end{document}